\documentclass[10pt,twocolumn,oneside]{pnas-new}
\usepackage{flushend}
\usepackage{adjustbox}
\usepackage{dcolumn}
\usepackage{longtable}
\usepackage{booktabs}
\usepackage{lscape}
\usepackage{placeins}
\usepackage[defaultcolor=red]{changes}
\usepackage{textcomp}
\usepackage{tcolorbox}
\usepackage{subcaption}
\usepackage{bm}
\usepackage{multicol}
\usepackage{siunitx}

\usepackage{tikz}
\usetikzlibrary{arrows.meta,positioning, calc}

\usepackage[page]{appendix}
\renewcommand{\appendixpagename}{
  \begin{center}
    \begin{tabular}{c}
      \textsf{\huge Supplementary Information for} \vspace{1em}\\ 
      \textsf{\Large Coach not crutch:}\\
      \large Evidence that AI can improve writing skill despite reducing effort\\
    \end{tabular}
  \end{center}
}

\everymath={\sf}

\DeclareMathVersion{nxbold} 
\SetSymbolFont{operators}{nxbold}{OT1}{cmr} {b}{n}
\SetSymbolFont{letters}  {nxbold}{OML}{cmm} {b}{it}
\SetSymbolFont{symbols}  {nxbold}{OMS}{cmsy}{b}{n}

\templatetype{pnasresearcharticle} 
\title{Coach not crutch: Evidence that AI can improve writing skill despite reducing effort}
\author[1]{Benjamin Lira}
\author[2]{Todd Rogers}
\author[3]{Daniel G. Goldstein}
\author[1]{Lyle Ungar}
\author[1]{Angela L. Duckworth}
\affil[1]{University of Pennsylvania}
\affil[2]{Harvard University}
\affil[3]{Microsoft Research}

\leadauthor{Lira} 

\acknow{\footnotesize We thank the Wharton Behavioral Lab for their generous assistance with data collection and cost-sharing. Data, code, and materials are available \href{https://osf.io/tdcru/?view_only=8f5cd1393eac4268b2c372831f4e8cb0}{here}. We thank Hans Bode and Jacob P. for great research assistance. We thank Carmen Nobel and the Journalist Resource for assistance in recruiting professional editors, writers, and journalists. We extend special thanks to the many professional writers who generously contributed their time and expertise to this project, including Jim Davis, Theodore Fernando, Michael Ambrosino, Ronnie Willis, Jo Ann Livingston, Deborah Gdula, Mike Sherry, Jocelyn Dong, Kelly P. Kissel, Amabelle Ocampo, Katherine H. Donahue, Lindsay Pietenpol, Amanda Wolfe, Beth McLaughlin, Carol Katarsky, David Smith, Neil Reisner, Shawn Drury, Tanya Perez, Mary-Ann McHugh, Ashley Yeager, Pat Bellinghausen Zellar, Clarice Cardoso, Kathryn Marchocki, Simon C. Tung, Jones mark, Marion Davis, Christina Frangou, Nancy Peckenham, Stephanie Morrison, Mac McKerral, Mark Shwartz, Wacuka Mungai, Amanda Pearson, Elizabeth Blakey, Janet Raloff, Jim Johnson, Margaret Cadieux, Matt Wilson, Melanie Wachsman, Nettie Reynolds, Paul O'Donnell, Tim Wacker, Wendy Guild Swearngen, Nancy Lapid, Emily Matthews, Vladimire Herard, Steve Doig, and Lora Durance.}

\begin{abstract}
In a series of highly-powered empirical studies, we examine the intuition that by sparing effort, using AI inevitably hinders learning. First, in a nationally representative survey of young adults, the majority expressed the view that using AI makes people lazier and less capable. 
Next, in a random-assignment experiment, we gave participants a tutorial on best practices in professional writing, then provided one group with access to an AI writing tool and asked another to practice writing on their own. Those who practiced with AI indeed exerted less effort while practicing---yet wrote better cover letters in no-AI writing tests. In a second experiment with more rigorous control conditions, access to AI improved writing more than either googling cover letter examples and tips or receiving personalized feedback on their practice letters from experienced human editors. A third experiment explained these learning gains by showing that AI can teach by example: participants who viewed a cover letter that had been revised by the AI tool but did no further practice improved their writing as much as those who practiced writing with the original AI tool. Collectively, these pre-registered experiments suggest that AI can exert opposing effects on effort and learning rate---making it possible in at least some cases to work less and learn more.  
\end{abstract}

\dates{This manuscript was compiled on \today}

\begin{document}

\newcolumntype{.}{D{.}{.}{-1}}
\newcolumntype{s}{B{.}{.}{-2}}
\makeatletter

\keywords{\vspace{0.25em}\sf\textsc{Human Capital Development | Generative AI | Skill Acquisition}\vspace{1em}}

\makeatother

\maketitle
\thispagestyle{firststyle}
\ifthenelse{\boolean{shortarticle}}{\ifthenelse{\boolean{singlecolumn}}{\abscontentformatted}{\abscontent}}{}

Generative AI (henceforth AI) tools are increasingly powerful and prevalent \cite{bubeck2023}, and there is mounting evidence that they can dramatically boost performance. 
  For example, working side-by-side with AI as a copilot has been shown to increase both quality and speed in a variety of professional writing tasks (e.g., emails, memos, short reports) \cite{noy2023, dellacqua2023, wiles2023algorithmic}.

Nevertheless, advances in AI in recent years have been accompanied by concerns that immediate gains in performance are offset by the subsequent erosion of human capital~\cite{hofman2023sports, puntoni2021, borges2024}. For example, in a 2024 poll of American adults, 62\% predicted that generative AI would “lead to humans becoming less intelligent’’ \cite{hawkins2024between}. In a 2024 survey of K–12 educators, four times as many judged AI to be net harmful (24\%) as net beneficial (6\%) \cite{lin2024a}.

Concerns that AI hinders learning are justified for at least three reasons. 
First, AI sometimes confidently asserts erroneous facts (i.e., hallucinations \cite{ji2023survey, yuan2023well}).
Second, using AI may create an illusion of mastery \cite{fisher2015}, particularly if users conflate its skilled responses with their own. 
Third and most important, using AI may reduce human effort because the more AI solves your problems, the less you have to think about them yourself \cite{kosmyna2025your,bastani2024}. All things being equal, less effort results in less learning \cite{bjork2011making, james1983talks}.

Analogous deskilling effects have been documented with prior technologies. 
In cross-sectional and longitudinal studies, drivers who rely on GPS exhibit poorer hippocampal-dependent spatial memory \cite{maguire2000, griesbauer2022london}. Likewise, knowing that information can be easily retrieved from the internet reduces the likelihood that one will commit it to memory. \cite{sparrow2011}.
AI may be uniquely harmful because, unlike search engines or GPS, it can replace the process of thinking itself—not just the storage or retrieval of facts.

The assumption that diminished effort inevitably reduces learning is shaping the design of real-world AI learning tools.
For example, when Khan Academy released its AI tutor Khanmigo, it announced on its website that it would prevent students from seeing full solutions: “Khanmigo challenges you to think critically and solve problems \textit{without giving you direct answers}\footnote{emphasis added}” \cite{khanmigo2025}. 

\begin{figure}[ht]
\centering
\begin{tikzpicture}[
  node distance=10mm and 10mm, % <<< reduced horizontal distance
  box/.style={
    fill=gray!20,
    minimum width=22mm,       % <<< smaller box width
    minimum height=10mm,
    align=center,
    font=\fontfamily{phv}\selectfont
  },
  >={Stealth[length=2.4mm]},
  every path/.style={thick}
]

% Left (AI Use)
\node[box] (ai) {AI Use};

% Center mediators (stacked)
\node[box, right=of ai, yshift=7mm] (effort) {Effort};
\node[box, right=of ai, yshift=-7mm] (utility) {Learning\\Environment};

% Midpoint between effort and utility
\coordinate (mid) at ($(effort)!0.5!(utility)$);

% Right (Skill) vertically centered
\node[box, right=20mm of mid] (skill) {Skill};

% Arrows AI -> Mediators
\draw[->] (ai) -- (effort);
\draw[->] (ai) -- (utility);

% Arrows Mediators -> Skill
\draw[->] (effort) -- (skill);
\draw[->] (utility) -- (skill);

\end{tikzpicture}

\caption{\textbf{Conceptual framework.} Relying on AI may cause people to exert less effort but improve the learning environment. Post-practice writing skill depends on the net effect of effort and the quality of the learning environment.}
\label{fig:arrows}
\end{figure}

Skill acquisition, however, depends on more than effort; learning rate also matters \cite{duckworth2015mechanics}.
A wide array of factors have been shown to influence how quickly students learn in conventional educational settings \cite{wisniewski_power_2020,bloom_2_1984, sana_interleaving_2022}.
One feature of optimal learning environments with special relevance to the affordances of generative AI is the availability of specific examples that illustrate abstract principles that are otherwise difficult for the learner to grasp and apply. 
Separate from its influence on effort, therefore, AI might change the quality of the learning environment. See Figure \ref{fig:arrows}.
The net effect of effort and learning environment on skill acquisition remains an open empirical question.

Here, we focus on writing---the most common use of AI at work. 
In a nationally representative sample of American adults in 2024, writing tasks 
were the most frequent workplace application \cite{bick2024rapid}. 
This pattern is corroborated by large-scale usage data: In an analysis of over 1 million 
ChatGPT conversations writing constituted 
24\% of all messages and 40\% of work-related messages, making it the dominant 
occupational use case \cite{chatterji_how_nodate}. Notably, approximately two-thirds of writing-related 
queries involved modifying existing text rather than generating novel content.

We know of only one study \cite{kosmyna2025your} that has experimentally manipulated access to AI (vs. no access) during a writing task. Fifty-four undergraduates at four selective universities were randomly assigned to condition and then asked to write essays: 18 participants wrote with access to AI, 18 wrote without access to AI, and 18 wrote with access to Google Search. Although a no-AI test of writing skill was not administered, compared to the other two conditions, participants given access to AI were less likely to recall a verbatim sentence from their essays immediately after writing them.

\begin{figure*}[t]
    \centering
    \includegraphics[width=.9\linewidth]{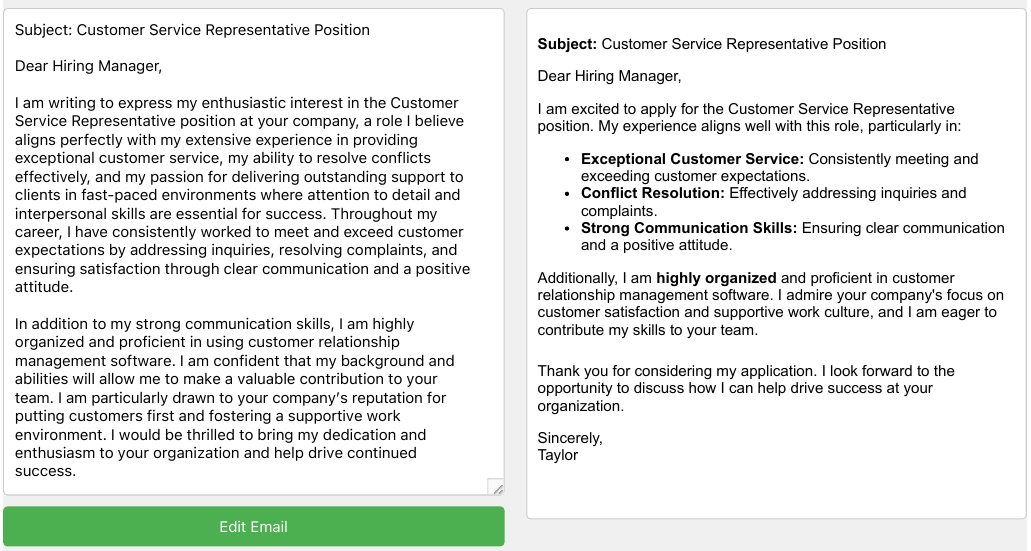}
    \caption{The AI writing tool we created for this investigation mimics a widely used and publicly available AI writing tool that takes input text (left panel) and generates a version that incorporates recommended writing principles (right panel).}
    \label{fig:tool}
\end{figure*}

\begin{figure*}[]
    \centering
    \includegraphics[width=.9\linewidth]{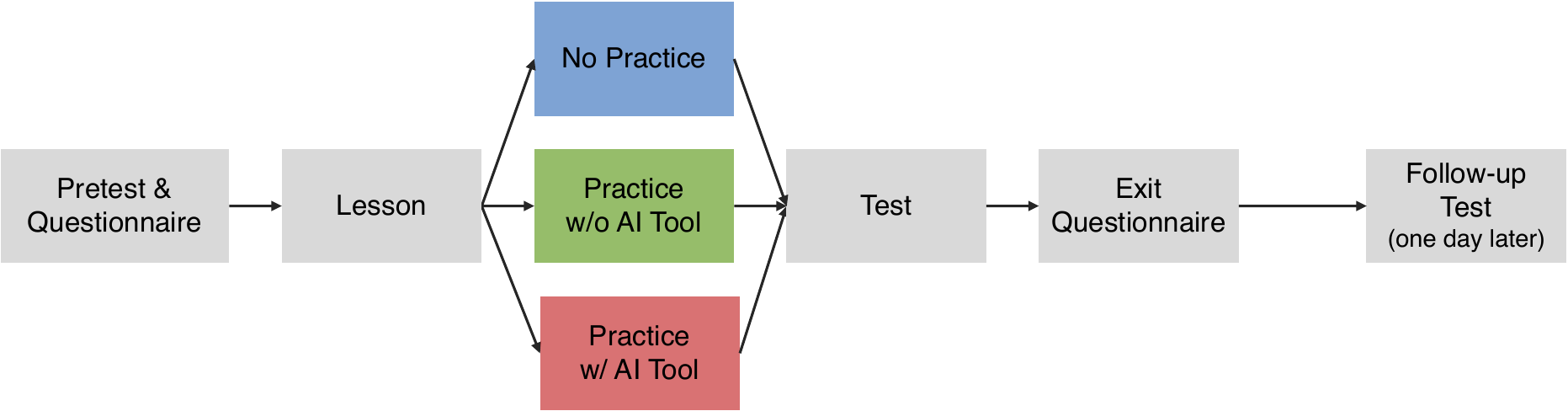}
    \caption{\textbf{Experimental design for Study 2.} First, all participants completed a baseline questionnaire, a pretest (rewriting a poorly-written cover letter), and a lesson introducing five evidence-based principles of effective writing \cite{rogers2023}. Next, participants were randomly assigned to one of three conditions: practicing with an AI writing tool, practicing without an AI writing tool, or no practice. Then, all participants were tested on writing skill (rewriting a new cover letter without access to AI) and completed an exit questionnaire. Finally, to assess the persistence of skill improvement, participants were invited to complete a similar incentivized test of writing skill one day later.}
    \label{fig:design}
\end{figure*}

\begin{figure}
    \centering
    \includegraphics[width=\linewidth]{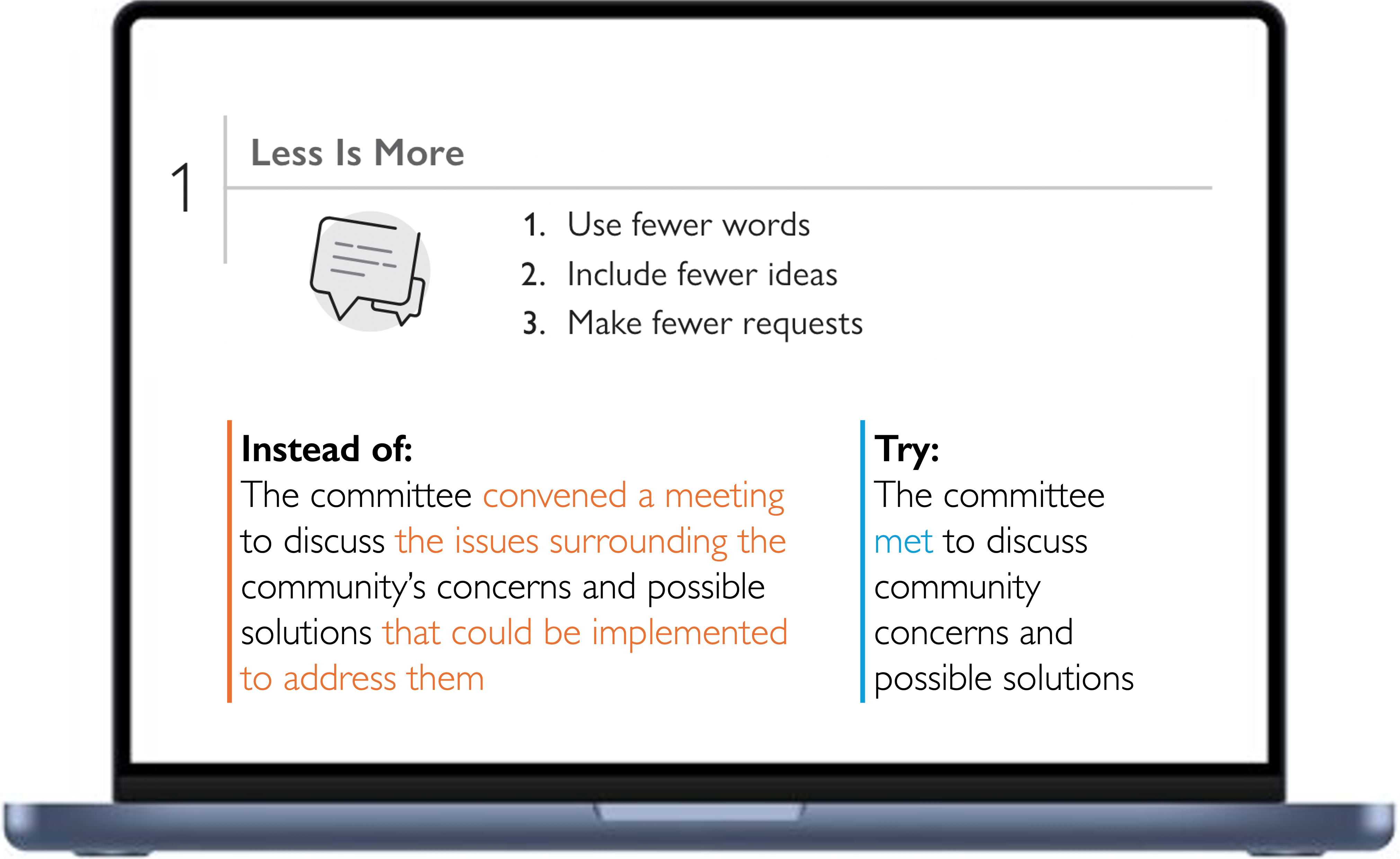}
    \caption{Screenshot of one of the pages of the lesson shown to all participants.}
    \label{fig:tutorial}
\end{figure}

We developed an experimental paradigm to determine whether using AI as a productivity tool for writing cover letters enhances or erodes subsequent writing skill. To begin, all participants complete a baseline questionnaire and writing pretest (i.e., revising a poorly written cover letter). Next, as shown in Figure \ref{fig:tutorial}, all participants are taught five evidence-based principles for professional writing (i.e., Less is more, Make reading easy, Design for easy navigation, Use enough formatting but no more, Make responding easy) \cite{rogers2023,shulman2024} and are shown a domain-general example for each principle. 
After this lesson, participants are randomly assigned to different types of practice (e.g., practice with access to an AI tool, practice without access to an AI tool, practice with feedback from professional editors, and practice with Google Search to find examples and tips). 
Finally, all participants rewrite a new cover letter \textit{without access to AI} (and with copy-pasting functionality disabled). 
Each participant's submitted cover letter is rated on adherence to each of the five writing principles, respectively. To gauge the ecological validity of the resulting composite index, we recruited a separate sample of raters (blind to condition) who read randomly chosen pairs of cover letters from the main experiment and were asked to ``choose the cover letter that would make you more likely to offer an interview to the candidate.'' See Methods for details.

% MOVE TO METHODS: Afterward, we evaluate the quality of participants' writing in two ways. First, we used AI to assess the adherence to each of the taught strategies on an 11-point scale. We averaged these ratings to produce a summary score (Cronbach's $\alpha$ = .81). These ratings correlated with human ratings as much as a pair of trained human raters agreed with each other (\textit{r}$_{human-human}$ = .74, \textit{p} < .01, \textit{r}$_{human-AI}$ = .70, \textit{p} < .001, difference in rs, \textit{p} = .56, \textit{n} = 100). This measurement approach is conservative: participants practiced with GPT-4o but were evaluated by Claude Sonnet, and models tend to undervalue text from competing models (i.e., same model bias, CITE). We establish the ecological validity of the task by asking a separate sample of \textit{human} evaluators naive to this investigation to read two letters and select the one more likely to secure its writer a job interview.
% These two operationalizations converged: preferred cover letters obtained higher AI ratings of writing skill (\textit{rs}~=~.29, \textit{p}s < .001). 

% INCREASE DISTANCE FROM INTERVENTION TO DV.

In this investigation, we examine contemporary assumptions about learning and AI and empirically test the accuracy of these beliefs. In Study 1, a nationally representative survey of young adults in the U.S. in October 2025, the majority endorsed the view that using AI makes people lazier, less smart, and less educated.
In Study 2, participants who practiced writing cover letters with AI actually learned more than participants who practiced without AI---an advantage that persisted in a one-day follow-up test. A separate sample of lay forecasters predicted the opposite result.
In Study 3, again against forecaster predictions, participants who practiced with AI learned more than participants who practiced with feedback from professional editors (or those who practiced with Google Search to find examples and tips).  
Finally, in Study 4, we explored the mechanism for these learning gains by introducing an example-only condition. 
Participants who viewed an AI-generated example (but did not have an opportunity to edit it) improved their writing as much as participants who practiced with AI; benefits again persisted in a one-day follow-up test. 

\section*{Study 1: Public perceptions of AI and learning}

In partnership with the Gallup Organization, we asked a nationally representative sample of American adults aged 18 to 28; (\textit{N} = 2,637) in October 2025 about AI use in the past month and attitudes about AI in general. 

Consistent with 2024 polls, writing is the most common use case for AI (45\% of all adults surveyed) after using AI as a replacement for Google (64\%) and ahead of generating images (35\%) or help with math (23\%). In general, AI is used more commonly for work (53\%) than for personal reasons like life advice (28\%) or companionship (17\%). 

% Using a 6-point Likert-type scale, (1 = Strongly disagree, 6 = Strongly agree), 
Participants expressed concern that ``AI makes people lazier'' (79\%, One sample \textit{d} = 0.66, \textit{p} < .001) and, to a more moderate degree, that ``AI makes people learn less'' (62\%, One sample \textit{d} = 0.34, \textit{p} < .001) and ``AI makes people less smart'' (63\%, One sample \textit{d} = 0.30, \textit{p} < .001). See Figure \ref{fig:gallup}. 
See Section \ref{sec:gallup} in the Supplementary Information for details.

\begin{figure}
    \centering
    \includegraphics[width=1.1\linewidth]{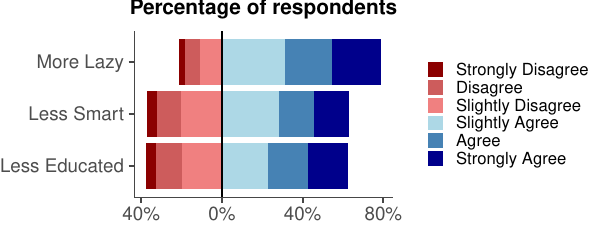}
    \caption{In a nationally representative sample of young adults, most endorsed the view that AI makes people less smart, lazier, and less likely to learn.}
    \label{fig:gallup}
\end{figure}

\section*{Study 2: Effect of practicing with AI on learning} 

% MAIN LINES
\begin{figure*}[t]
    \centering
    \includegraphics[width=1\linewidth]{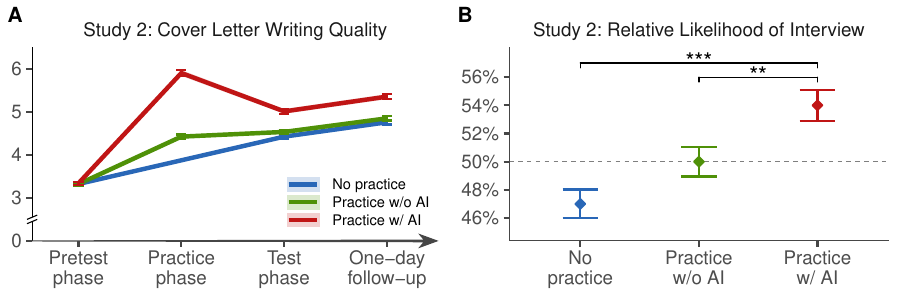}
    \caption{\textbf{A. Cover Letter Writing Quality,} In both the main test and the follow-up of Study 2, participants who had practiced with AI outperformed those who practiced without it and those who did not practice at all.
    Error bars represent means $\pm$ 1 SE.
    Means shown are for the subsample of participants (\textit{n} = 1,294) who completed the one-day follow-up test. See Figure \ref{fig:s2_supplement} for the equivalent figure in the full sample. \textbf{B. Relative Likelihood of Interview.} During the test phase of Study 2, cover letters written by participants who had practiced with AI were judged more likely to generate hypothetical interview offers than those from other conditions. Error bars represent proportions $\pm$ 1 SE.}
    \label{fig:s2_combined}
\end{figure*}

% Study 3 tested whether the predictions of Study 1 forecasters were
% accurate. Specifically, 
% In Study 2, \textit{N} = 2,238 participants completed a baseline questionnaire and pretest (rewriting a poorly written cover letter), followed by a lesson introducing five principles of effective professional writing.
In Study 2 (\textit{N} = 2,238), we preregistered an experiment in which participants completed a baseline questionnaire and pretest, followed by a lesson on effective writing. They were then randomly assigned to one of three practice conditions: (1) rewriting a new cover letter with an AI writing tool that instantly revised text based on these principles, (2) rewriting the new cover letter without the AI tool, or (3) a no-practice control. 
Because forgetting occurs rapidly—with substantial decay even within 24 hours (e.g., \cite{roediger_test-enhanced_2006, murre2015replication}), we invited all participants to complete a similar incentivized test of writing skill one day later. See Figure \ref{fig:design}.

\subsection*{AI practice improved writing skill}

% Consistent with other studies demonstrating the productivity benefits of
% AI tools \cite{dellacqua2023, noy2023}, participants given access to the
% AI writing tool produced cover letters during the practice phase that
% were dramatically higher in quality than participants without access
% (\emph{d} = 1.01, \emph{p} \textless .001).

Consistent with our preregistered hypothesis, participants who had practiced with the AI tool wrote better cover letters than
participants who either had practiced without the AI tool (\emph{d} = .38,
\emph{p} \textless .001) or who had not practiced at all (\emph{d} = .47,
\emph{p} \textless .001). See Figure \ref{fig:s2_supplement}.
Likewise, cover letters written by participants who had practiced with AI were more likely to secure a hypothetical job interview than cover letters by participants who had practiced without AI (.54 vs. .50, \textit{p} = .002) or had not practiced at all (.54 vs. .47, \textit{p} < .001). See Figure \ref{fig:s2_combined}B. %Notably, the effect of practicing with AI was not limited to stronger or weaker writers---participants across all skill levels benefited equally from practicing with AI. 

\subsection*{AI practice was less effortful}

Compared to participants who practiced alone, participants who practiced with the AI
tool spent 0.44 fewer minutes during the practice phase (3.73 vs.~4.17;
\emph{d} = $-$.12, \emph{p} = .025), logged roughly a quarter as many total
keystrokes (26\% \emph{d} = $-$.44, \emph{p} \textless .001) and logged 65\% fewer keystrokes per minute (\emph{d}~=~$-$.22, \emph{p} \textless .001),  and
self-reported expending less effort during practice (\emph{d~}~=~$-$.31,
\emph{p} \textless .001).

Nevertheless, it would be inaccurate to label writers practicing with AI as entirely disengaged. 
Rather than simply copying and pasting the AI tool's output, the majority of participants practicing with AI chose to interact with the task for over 3 minutes, and 95\% made at least one edit to the AI-generated output before final submission. See Figure \ref{fig:distance} in Supplementary Information for details.

% Differential effort during practice raised the possibility that
% participants who had practiced with the AI tool outperformed those who
% had practiced on their own because they were less fatigued during the test phase. During the test phase, however, participants who had practiced with AI spent about as much time as those who had practiced without it (\emph{d} = .06, \emph{p~}~=~.235), but logged more keystrokes (\emph{d} = .12, \emph{p} = .026), and self-reported expending less effort (\emph{d} = $-$.12, \emph{p} = .023). ADD KPM

\subsection*{AI practice did not create the illusion of mastery}
% Despite improving more in objectively assessed writing skill, participants who had practiced with AI did not overestimate their learning. 
In the exit questionnaire, participants who had practiced with AI  reported learning about as much as participants who had practiced without it or not practiced. 
Likewise, they judged their writing skill similarly to those who had either practiced alone or done no practice at all (|\textit{d}s| $\le$ 0.10, \textit{p}s \textgreater .05).
When asked if they wanted to see additional feedback, they were about as likely to say yes as participants who had practiced without AI (.64 vs.~.60, \textit{p~}~=~.167), and slightly less likely than those who had not practiced at all (.65 vs.~.60, \textit{p~}~=~.039).
See Section \ref{sec:future_learning2} in the Supplementary Information
for details.

% Participants who had practiced with AI reported having learned about as much and judged their writing skill similarly to those who had either practiced alone or done no practice at all (\textit{p}s \textgreater .05).
% When asked if they wanted to see optional feedback after the test, they were about as likely to say yes as participants who had practiced without AI (.64 vs.~.60, \textit{p~}~=~.167), and slighlty slightly less likely than participants who did not practice (.65 vs.~.60, \textit{p~}~=~.039).

% Finally, participants who had practiced with AI were slightly less likely to request feedback after
% the test phase (.65 vs.~.60, \textit{p~}~=~.039), but just as likely as
% participants who had practiced without AI (.64 vs.~.60, \textit{p~}~=~.167).
% 
\subsection*{The benefits of practicing with AI were just as large one day later}
To examine whether the treatment effects persisted over time, we re-contacted all
participants one day later. Most participants responded
(87\%), and attrition rates did not differ by condition (13\% to
14\% \(\chi^2\)~=~.68, \textit{p} = .710). Confirming our pre-registered
hypothesis, participants who had practiced with the AI tool the
previous day continued to outperform those who had practiced without the
tool (\emph{d} = .41, \emph{p} \textless .001) as well as those who had
not practiced at all (\emph{d} = .46, \emph{p~}\textless~.001).
Participants who had practiced without the AI tool performed no better
than those who did not practice (\emph{d} = .05, \emph{p~}~=~.331). 
The magnitude of the follow-up effect size was not significantly different from the immediate test phase effect size (condition $\times$ time interactions were non-significant (\textit{p}s > .69).
See Figure \ref{fig:s2_combined}A and Section \ref{sec:persists2} of Supplementary Information for details.

\subsection*{Results were not moderated by individual differences}
None of the findings above were moderated by individual difference
variables, including past experience with AI, age, gender, race,
education, motivation to learn, and baseline writing skill, BH-corrected
\textit{p}-values \textgreater .05. See Table \ref{tab:interactions2} in
the Supplementary Information for details.

\subsection*{Forecasters predicted AI would reduce effort and harm learning}
We showed \textit{N} = 150 participants screenshots of this design and asked them to rank-order the study conditions according to how much they predicted participants would learn in each. 
Confirming our pre-registered hypothesis, nearly twice as many forecasters (64.7\%) ranked practicing alone above practicing writing with access to an AI tool as the converse (35.3\%, $\chi^2$ (1) = 12.9, \textit{p} < .001), see Figure \ref{fig:s1}. 
Participants made this prediction regardless of self-reported experience with AI (\textit{OR} = 0.72, \textit{p} = .096) or any other measured demographic characteristic (\textit{p}s > .05). See Table \ref{tab:ORs1}

Forecasters who were pessimistic about the effect of the AI tool on learning speculated that it would crowd out effort (e.g., ``Practicing alone would force more recall and problem-solving skills, while AI essentially just gives them the answer.'', ``I think oftentimes using AI impedes the learning process because it's the `easy way.'''). Those with positive views, on the other hand, cited the possibility of AI providing insights or examples that would be otherwise unavailable (``As much as I hate AI, I do not believe you can improve in any manner if you do not have examples or other ways of learning, and AI can provide this.'')

\section*{Study 3: Practicing with AI vs. with expert feedback (vs. with Google Search)}

In Study 3 (\textit{N} = 2,997), we preregistered an experiment testing whether AI was more helpful than two ecologically valid comparison conditions.
Participants (\textit{N} = 2,997) completed a baseline questionnaire and pretest, followed by the lesson on effective writing, and then practiced rewriting a new cover letter on their own. Next, participants were randomly assigned to one of three conditions: (1) practice with AI, (2) practice with feedback from professional human editors, (3) practice with Google Search. Letters written by participants assigned to the second condition were given to professional editors averaging 25 years of professional editorial and writing experience. These 49 editors, recruited in partnership with the Journalists' Resource at Harvard Kennedy School, provided personalized revisions and constructive feedback. The following day, all participants attempted to improve the cover letter they had previously submitted, using the resources afforded by their assigned condition. Finally, all participants completed a writing test (rewriting a cover letter without access to AI) and an exit questionnaire.

\subsection*{Practicing with AI improved writing skill more than feedback from professional editors and Google Search}

During the practice phase, participants working with AI outperformed participants who received personalized feedback from professional editors (\textit{d} = 0.76, \textit{p} \textless .001) and those assigned to Google cover letter examples and tips (\textit{d} = 1.03, \textit{p} \textless .001).

\begin{figure*}[tb]
    \centering
    \includegraphics[width=1\linewidth]{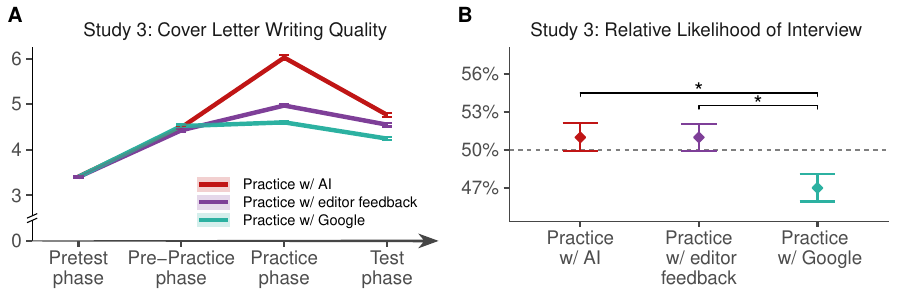}
    \caption{\textbf{A. Cover Letter Writing Quality.} In the test phase, Study 4 participants who had practiced with the AI tool outperformed those who had practiced with either feedback from professional editors or access to Google Search.
    Error bars represent means $\pm$ 1 SE. \textbf{B. Relative Likelihood of Interview.} In the test phase, cover letters written by participants who had practiced with AI were about as likely to generate hypothetical interview offers as those written by participants who had practiced with feedback from professional editors. Cover letters from both of these conditions outperformed those written by participants who had practiced with Google Search. Error bars represent proportions $\pm$ 1 SE.}
    \label{fig:s5_combined}
\end{figure*}

During the test phase, in which participants were asked to write a new cover letter without access to any outside resources, participants who had practiced with the AI tool wrote higher-quality cover letters than those who had practiced with Google Search (\emph{d} = 0.46, \textit{p} \textless .001), and even outperformed those who had practiced with personalized feedback from professional editors (\textit{d} = 0.20, \emph{p} \textless .001). See Figure \ref{fig:s5_combined}A. 

Cover letters written by participants who had practiced with AI were as likely to secure a hypothetical job interview as cover letters by participants who had practiced with editor feedback (.51 vs. .51, \textit{p} = .843), and more likely to than those written by participants who had practiced with Google Search (.51 vs. .47, \textit{p} = .024). See Figure \ref{fig:s5_combined}B.

% HUMAN  BENCHMARKS
% spending less time per letter as the task progressed (\textit{b} = -0.387, SE = 0.028, \textit{p} < .001), the quality of their edits and feedback actually improved (\textit{b} = 0.013, SE = 0.005 \textit{p} < 0.001).

A possible concern is that reviewing many letters consecutively caused editors to get tired and give worse feedback over time. This could suggest that our experiment compared AI against a  ``strawman'' version of human expertise. However, analysis of the editors’ performance revealed the opposite pattern: while editors became more efficient, spending less time per letter as the task progressed (\textit{b} = $-$0.387, \textit{p} < .001), the quality of their edits and feedback actually improved (\textit{b} = 0.013, \textit{p} < 0.001). Thus, the AI tool’s advantage over professional editors cannot be attributed to a decline in editorial performance. See Section \ref{sec:editors} of Supplementary Information for details.

The quality of edits provided by AI was higher than that of edits from professional human editors. We used the AI tool to rewrite each of the texts assigned to the editor pool, creating a counterfactual of how AI would have edited those letters. Relative to the counterfactual, letters rewritten with the AI tool were rated higher for writing quality (\textit{d} = 1.66, \textit{p} < .001), and when shown to independent raters, they were considerably more likely to secure a hypothetical job interview (\textit{d} = 0.75, \textit{p} < .001). 
Thus, AI users encountered examples that more closely adhered to the writing principles than those provided by human editors, helping explain the learning advantage associated with AI use.

See Section \ref{sec:editors} of Supplementary Information and Figure \ref{fig:ai_vs_editors} for details.

\subsection*{Practicing with AI was not more effortful than getting feedback from professional editors and was less effortful than looking for examples online}

During the practice phase, participants with access to AI spent about the same time as those who practiced with editor feedback (difference = 0.43 minutes, \emph{d} = 0.09, \textit{p} > .05), but about 1.18 fewer minutes than those who practiced with Google Search (\emph{d~}=~0.32, \emph{p~}<~.001). Nevertheless, participants across all three groups reported similar levels of subjective effort (\emph{p}s~>~.05). Likewise, participants who practiced with AI logged about as many keystrokes as those who practiced with editor feedback (68\% fewer, \emph{d} = 0.08, \textit{p} > .05) and 76\% fewer keystrokes than those who practiced with Google Search (\emph{d~}=~0.32, \emph{p} < .001). They also typed at a slower rate than participants who practiced with editor feedback (\emph{d~}= $-$0.41, \emph{p}~<~.001) or those who practiced with Google Search (\emph{d}~=~$-$0.34, \emph{p} < .001).

% In the test phase, differences in effort narrowed. Participants who had practiced with AI spent about the same as those who had practiced with editor feedback (median = 5.72 vs.\ 5.52 minutes, \emph{d} = 0.07, \textit{p} > .05) and slightly more time writing than participants who had practiced with Google Search (median = 5.72 vs.\ 5.16 minutes, \emph{d} = 0.18, \emph{p} < .001). Self-reported effort did not differ between participants who had practiced with AI and participants who had practiced with editor feedback (\textit{d} = .02, \emph{p} > .05), though participants who had practiced with Google Search rated the task as easier than those who had practiced with AI (\emph{d} = $-$0.22, \emph{p} < .001) and participants who received editor feedback (\emph{d}~=~$-$0.24, \emph{p} < .001). Participants across the three conditions accrued approximately the same amount of keystrokes (\emph{p}s~>~.05).

\subsection*{No evidence of AI-induced illusion of mastery}

Participants who had practiced with AI did not report learning more than those who had practiced with editor feedback (\textit{d} = 0.07, \textit{p} > .05), nor did they consider their writing skills to be better (\textit{d}~=~$-$0.02, \textit{p}~>~.05). Both groups, however, correctly recognized that they had learned more than those who had practiced with Google Search (\textit{d}s = .28 and .20, respectively), and evaluated their writing skills accordingly (\textit{d}s~=~.11 and .12, respectively). Participants who had practiced with AI and those who had practiced with editor feedback requested feedback on their test submissions at similar rates (71\% and 70\%, respectively). Both groups were more likely to request feedback than those who had practiced with Google search (64\%, \textit{p} = .003 and .007, respectively). See Section \ref{sec:future_learning5} in the Supplementary Information for details.

\subsection*{Results were not moderated by individual differences}
As in Study 2, the above findings were consistent across all measured individual differences. See Table \ref{tab:interactions5} in the
Supplementary Information for details.

\subsection*{Forecasters predicted AI would be less helpful than human editors and were less willing to pay for it}
We asked \textit{N} = 150 people to imagine they were participants in our experiment and asked them how much they would be willing to pay for feedback from the AI tool vs. feedback from experienced human editors. 
Out of a maximum of a dollar (the total amount of the hypothetical bonus participants could earn), participants on average were willing to pay 28 cents for the AI tool vs. 38 cents for feedback from an experienced human editor (\textit{t}(149)~=~$-$4.83, \textit{p}~\textless~.001, \textit{d}~=~0.39). 
Participants predicted that experienced human editors (\textit{M}~=~6.68 on a 0---10 Likert scale) would be more effective in helping them improve their writing, compared to feedback from an AI tool (\textit{M}~=~5.64; \textit{t}(148)~=~$-$4.77, \textit{p}~\textless~.001, \textit{d}~=~0.39). People's beliefs about the effectiveness of AI and human editor feedback predicted willingness to pay (b = 0.06, SE = 0.01, \textit{p} < .001) and accounted for most of the preference for human editors (indirect effect = 0.06, \textit{p} < .001). Once these predictions were accounted for, the remaining preference for human feedback was reduced by 60\% (c' = 0.04, SE = 0.02, \textit{p~}=~.054). See Figure \ref{fig:mediation} and Section \ref{sec:mediation} of Supplementary Information for details.

\section*{Study 4: Examples as the mechanism for AI learning gains} 

% MECH COMBINED
\begin{figure*}[t]
    \centering
    \includegraphics[width=\linewidth]{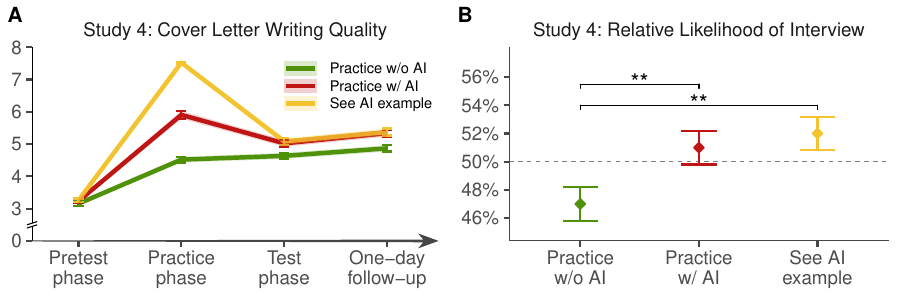}
    \caption{
    \textbf{A. Cover Letter Writing Quality.} In both the main test and the follow-up of Study 4, participants who simply saw an AI-generated example improved just as much as those who practiced with AI and more than those who practiced without AI.
    Error bars represent means $\pm$ 1 SE.
    Means shown are for the subsample of participants  (\textit{n} = 608) who completed the one-day follow-up test.
    See Figure \ref{fig:s3sup} for the equivalent figure in the full sample. \textbf{B. Relative Likelihood of Interview.} During the test phase, cover letters written by participants who had seen an AI-generated example were about equally likely to generate hypothetical interview offers when compared to those assigned to practice with AI. Cover letters from both AI conditions outperformed those written by participants assigned to practice without AI. Error bars represent proportions $\pm$ 1 SE.}
    \label{fig:s3_combined}
\end{figure*}

To better understand how using AI improves writing skill, in Study 4 (\emph{N} = 2,003), we preregistered a replication and extension in which we replaced the no-practice condition of Study 2 with an example-only condition.
In this condition, we showed participants an AI-generated writing example that they could not edit. 
To the extent that the benefit of practicing with AI was driven by exposure to examples, the example-only condition should improve performance in the test phase as much as the practice with AI condition.

\subsection*{Seeing an AI example was as effective as practicing with
AI}\label{seeing-an-ai-example-was-as-effective-as-practicing-with-ai}

As in Study 2, participants given access to the AI writing tool dramatically outperformed participants who did not get access to it, both while using it during the practice phase (\emph{d} = 1.22, \emph{p} \textless .001), and during the no-AI test phase (\emph{d} = .34, \emph{p}  \textless{} .001). Their test phase cover letters were also relatively more likely to secure them hypothetical job interviews (.51 vs. .47, \textit{p} = .008).

Participants who had merely seen an AI-generated example also improved more in writing skill than those who had practiced without AI (\emph{d} = .37, \emph{p} \textless{} .001), and produced letters that were relatively more likely to secure them interviews (.52 vs. .47, \textit{p} = .007). Notably, they improved as much as participants who had practiced with the AI tool (and could edit its output, \emph{d} = .03, \emph{p} = .883), and were offered hypothetical interviews at similar rates as them (.51 vs. .52, \textit{p} = .561). See Figures \ref{fig:s3sup} and \ref{fig:s3_combined}B. %As in Study 2, gains from practicing with AI and seeing AI examples were comparable across all levels of baseline writing skill.

\subsection*{Seeing an AI example was even less effortful than practicing
with
AI}\label{seeing-an-ai-example-was-even-less-effortful-than-practicing-with-ai}

During the practice phase, participants who saw the AI example spent 2.32 fewer minutes than participants
practicing with AI (\emph{d} = 0.99, \emph{p} \textless{}
.001) and 2.96 fewer minutes than participants practicing without AI (\emph{d}~=~1.13 \emph{p} = \textless{} .001), and reported expending less effort than those practicing with an AI tool
(\emph{d} = 0.19, \emph{p} = .003) and those practicing without AI (\emph{d} = 0.32, \emph{p}
\textless{} .001).  As in Study 2, participants who practiced with AI logged 74\% fewer keystrokes and 68\%  fewer keystrokes per minute (\emph{d}~=~0.15, \emph{p} = .007) compared to participants who practiced without AI. As expected, participants exposed to the AI example logged 0 keystrokes.

\subsection*{Seeing an AI example did not create the illusion of mastery}

As in Studies 2 and 4, despite learning more, participants who had practiced with AI or had merely seen an AI-generated example reported learning similar amounts to those who practiced without AI (\textit{p}s~>~.05) and rated their writing skill after practice at comparable levels (\textit{p}s~>~.05). Moreover, all participants requested feedback at similar rates (proportions ranged from 63\% to 67\%). See Section \ref{sec:future_learning3} in the Supplementary Information for details.

\subsection*{The benefits of seeing an AI example were just as large a day later}\label{the-effects-of-seeing-an-ai-example-persist}

When we recontacted a subsample of participants (\textit{n}~=~800) one day later, 
  the majority responded (\textit{n}~=~633, 80\%); the attrition rates ranged from 17\% to 24\% and did not differ by condition ($\chi^2$ (2) = 4.56, \textit{p} = .102).
  The effect remained robust after 24 hours. Participants who had practiced with the AI tool (\emph{d} = .29, \emph{p} = .006) and participants who had simply seen an AI example (\emph{d}~=~.32, \emph{p} = .003), both continued to outperform those who had practiced without the tool. Participants who had merely seen an AI example performed as well as those who had practiced with AI (\emph{d} = .02, \emph{p} = .830). As in Study 2, the magnitude of the follow-up effect size was not significantly different from the immediate test phase effect size (condition $\times$ time interactions were non-significant (\textit{p}s > .96). See Figure \ref{fig:s3_combined}A and Section \ref{sec:persist3} of Supplementary Information for details.

\subsection*{Results were not moderated by individual differences}
As in the previous studies, the above findings were consistent across all measured individual differences. See Table \ref{tab:interactions3} in the
Supplementary Information for details.

\section*{Discussion} 

Although many believe that using AI undermines effort and learning, we found in our experiments that participants assigned to practice writing cover letters with AI learned more than those assigned to practice on their own. Access to AI also led to more learning than did access to Google search for cover letter tips and examples, or even personalized feedback from professional human editors. 

Lay beliefs about AI and effort were more accurate. Participants given access to an AI tool expended less effort while practicing than those who practiced alone or who practiced with access to Google tips and examples. Not surprisingly, 
practicing with AI was about as effortful as practicing with feedback from a professional human editor. 

What accounts for the learning gains from practicing with AI? 
We argue that AI improves the learning environment by providing high-quality, just-in-time, personalized examples that illustrate otherwise abstract principles. 
As shown in Figure \ref{fig:arrows}, even if AI reduces the effort people expend during practice, it may nevertheless improve learning by enhancing the learning environment. 
This mechanism is consistent with a large literature showing that humans are especially adept at learning by observation \cite{bandura1971, meltzoff2005imitation, lyons2007hidden}. 
Relatedly, individualized tutoring is a much more effective learning environment than whole group instruction \cite{bloom_2_1984}—and experts in diverse domains engage in deliberate practice, acquiring increasingly more sophisticated mental representations provided by their teachers and coaches \cite{anders2008deliberate}.
Compared to textbooks, conventional computer-tutoring programs (e.g., \cite{anderson1995cognitive}), and even human instructors, AI systems are especially adept at instantly generating specific, high-quality examples tailored to a learner’s momentary needs. 
Indeed, we found that AI-generated examples of cover letters during the practice phase were better written and more likely to elicit interviews than cover letters produced by professional writers and editors. 

Three promising directions for future research are worth highlighting.

First, it is not clear whether the observed benefits of AI examples would generalize to other domains like math \cite{kumar2023, bastani2024} and computer programming \cite{nie2024gpt}. 
At a glance, a single AI-generated example illustrates the principles of effective professional writing. 
In other domains, merely observing an example may be less informative. For instance, the final answer to a math or computer programming problem does not often reveal the process that produced it.

Second, additional research is needed to explore moderators. 
There may be metacognitive strategies that enhance the benefits of writing with AI. For instance, trying on one's own before using AI has been shown to improve learning more than the converse (cf. \cite{kumar2023}). 
Conversely, time pressure and competing priorities around performance and learning may minimize the learning benefits.  
More broadly, we believe concerns about AI as a crutch are justified when AI displaces effort without improving the learning experience.

Third, in our experimental paradigm, participants interacted with AI only once. 
It is common, however, to use AI repeatedly. 
  When do repeated interactions lead to diminishing or even negative returns, and in what scenarios might skill development continue over time?  Consider, for instance, the game of Go. The introduction of superhuman AI has been associated with an increase in the novelty and quality of decisions made by human Go players over time, with elite players reporting that they have been inspired by AI solutions they'd never seen before \cite{shin2023}.     
  Future research, ideally in field settings, is needed to establish the long-term benefits and costs of using AI.

Taken together, our findings challenge the widely held view that productivity boosts from AI inevitably hamper the development of human skill. 
By illustrating otherwise abstract principles with examples personalized to the precise needs of the learner at the moment, AI can, at least in some use cases, improve the quality of the learning environment. 
% AI produced better examples than professional editors, and of course, scaled better, and perhaps unexpectedly, led to \textit{more} learning. 
Rather than viewing AI as a universal threat to human capital, our results provide an existence proof that AI can support the development of skills at scale.

% Plausibly by enabling more learning with less effort, AI might, in the long-run, in fact encourage more cumulative practice over time.
% In some cases (which? thoughtful design, use, or particular use cases), AI tools can support both immediate performance and sustained learning outcomes that need not trade off against each other.
% This potential has not gone unnoticed. The emergence of ``learning-oriented'' modes in AI systems reflects demand from users who want tools that enhance their capabilities rather than replace them. These features suggest developers recognize that many users---programmers, writers, analysts---actively resist deskilling.

% The underappreciated efficacy of timely and tailored examples carries practical implications: 
% Many AI tools designed to support learning are explicitly programmed not to ``give away'' answers. 
% However, it may be that in addition to hints, leading questions, and explanations, learners benefit from demonstrations of the principles they are attempting to master.

\section*{Methods}
\subsection*{Ethical Considerations}
The study was assessed by the University of Pennsylvania's IRB, and was approved before implementation (Protocol 853653). All participants completed informed consent.

\subsection*{Pre-registration}
All studies were preregistered on \href{https://aspredicted.org}{https://aspredicted.org}. Accordingly, the analyses presented in the main text were also pre-registered, with exceptions noted below. 

Study 1 was a descriptive survey and was not preregistered.

Study 2 was pre-registered (\#201140). The pairwise comparisons analysis was pre-registered separately (\#205316). After preregistration, we considered a keystrokes-per-minute metric of effort intensity rather than duration; thus, this analysis was not preregistered. The forecasting data was preregistered (\#191800), but not the moderation analysis within it.

Study 3 was pre-registered (\#239157). The pairwise comparisons analysis (\#243115) and the comparison between editor rewrites and counterfactual AI rewrites (\#255792) were pre-registered separately. The keystrokes-per-minute metric of effort was not pre-registered. The willingness to pay data was preregistered (\#236487), but not the analyses on the predicted effectiveness of AI vs. human feedback and the mediation analysis. 

Study 4 was pre-registered (\#197704). The one-day follow-up was collected in three batches: a pilot, followed by second and third batches. We pre-registered batch 2 (\#199451), but report pooled results in the main text, with separate batch details available in Supplementary Information Section \ref{sec:persist3}. The pairwise comparisons analysis was pre-registered separately (\#205315). The keystrokes-per-minute metric of effort was not pre-registered.

See all preregistrations \href{https://lirabenjamin.github.io/learning-not-cheating/all_preregs.pdf}{here}.
    
\subsection*{Participants}
Participants in Study 1 were a nationally representative sample of young adults aged 18-28 (M = 25.3, SD = 2.4) recruited through the Gallup organization. The sample comprised 38\% men, 57\% women, and 5\% nonbinary participants. The majority (64\%) were White, with the remainder comprising Hispanic (15\%), Asian (10\%), Black (10\%), and other race (1\%) participants. We used sample weights to ensure national representativeness of the data for all analyses.

We sampled participants from Prolific for all our experiments. We excluded all Prolific users who participated in one of the earlier studies from participation in subsequent studies. 

In Study 2, the sample was somewhat evenly split between men (46\%) and women (52\%), ranging in age from 18 to 82 (M = 36.0, SD = 12.5). Over half of the sample (58\%) was White, with the rest comprising Black (33\%), Latino (6\%), and Asian (5\%) participants. Most participants (77\%) had college degrees. Most participants (93\%) were at least somewhat motivated to improve their writing, and had varying levels of experience with AI writing assistants (36\% had tried them, but hardly ever used them, 47\% used them at least a few times per week, and 17\% had never used AI assistants before). Participants forecasting the results of Study 2 were predominantly female (\textit{n}~=~93, 62\%), with ages ranging from 21 to 81 (M~=~38.4, SD = 12.2). They were predominantly white (75\%). A small proportion were students (13\%), and most were employed (62\%).

In Study 3, the sample consisted of relatively fewer men (38\%) than women (61\%), and ranged in age from 19 to 92 (M = 41.5, SD = 13.8). Over half of the sample (74\%) was White, with the rest comprising Black (16\%), Latino (8\%), and Asian (7\%) participants. Most participants (74\%) had college degrees. Most participants (94\%) were at least somewhat motivated to improve their writing, and had varying levels of experience with AI writing assistants: 17\% had never used AI assistants, 40\% had tried them but hardly ever used them, and 43\% used them at least weekly. Participants forecasting Study 3 were evenly split between men (52\%) and women (48\%), and their ages ranged from 18 to 75 (M = 42.4, SD = 14.4). They were predominantly white (67\%). A small proportion were students (19\%), and most were employed (56\%).

Study 4 had similar proportions of men (46\%) and women (53\%), and participants ranged in age from 18 to 95 (M = 37.9, SD = 12.6). Over half of the sample (64\%) was White, with the remainder comprising Black (24\%), Latino (8\%), and Asian (6\%) participants. Most participants (74\%) had college degrees. Most participants (91\%) were at least somewhat motivated to improve their writing, and had varying levels of experience with AI writing assistants (40\% had tried them, but hardly ever used them, 42\% used them at least a few times per week, and 18\% had never used AI assistants before).

\subsection*{Procedure}

In Study 1, participants completed a brief survey about their AI usage in the past month and their general attitudes toward AI and learning. Participants rated their agreement with statements about AI's impact on learning, intelligence, and work habits using 6-point Likert-type scales (1 = Strongly disagree, 6 = Strongly agree).

All experiments (Studies 2--4) followed the same general structure with study-specific variations. Participants first saw an introductory screen and completed a brief questionnaire reporting demographics, prior experience with AI writing tools, motivation to improve their writing, and perceived writing skill. They then completed a short pre-test (2 minutes) in which they edited a poorly written cover letter, followed by a brief lesson on the five principles of effective writing.

In Studies 2 and 4, participants were then randomized to a practice condition (or skipped practice in the no-practice control). During practice, participants rewrote a new cover letter or, in the example-only condition, viewed an AI-rewritten version of the letter. This example was not explicitly labeled as AI-generated. After practice, participants saw either their own rewritten letter or the AI-generated example, along with AI-generated feedback highlighting one suggested improvement (see Supplementary Information, Section~\ref{sec:feedback}). Immediately after this feedback, participants rated how much they had learned and how hard they had worked so far. They then completed a 7-minute incentivized test of writing skill by editing a new cover letter without access to AI tools. To minimize the possibility of cheating, we used custom JavaScript to restrict copy-pasting functionality. Finally, participants could request optional feedback and were asked whether they had used outside resources during the test. A small percentage of participants (2.9\%) admitted to doing so; as per our preregistration, these participants are included in analyses, though results excluding them are consistent with our main interpretation (see Tables~\ref{tab:s2_test}, \ref{tab:test_effects_s5}, and \ref{tab:s3_test}).

In Study 3, we adapted the procedure to compare AI practice with two ecologically valid human alternatives: feedback from professional editors and guidance from Google Search. Participants completed the same baseline questionnaire, pretest, and writing lesson. They then edited a new cover letter without any assistance and were randomly assigned to one of three conditions: (1) practice with AI, (2) practice with feedback from professional editors, (3) practice with a Google Search for cover letter examples and tips. In the second condition, participants’ cover letters were forwarded to editors recruited in partnership with The Journalists’ Resource at Harvard Kennedy School, who provided both written feedback and a rewritten version of the letter. The following day, participants were asked to revise their original cover letter using the resources from their assigned condition. All participants then completed a final writing test by rewriting a new cover letter without assistance, followed by an exit questionnaire.

\subsection*{Measurement}
For Studies 2-4, as per our pre-registration, we used OpenAI's GPT-4o to rate text samples for writing quality. 
    To do this, we independently rated each cover letter and each writing principle. 
    Research has demonstrated that large language models can provide ratings of writing quality that align closely with human judgments, offering reliability and consistency across various evaluation contexts \cite{rathje2024, hackl2023}.
    See our prompts in Table \ref{tab:prompts}. 
    We then took the unweighted average of these 5 scores as our main dependent variable. 
    See disaggregated analyses by each writing principle and additional outcomes on Tables \ref{tab:s2_test}, \ref{tab:test_effects_s5}, and \ref{tab:s3_test}.

To validate these ratings, the first author and a trained research assistant took a sample of \textit{n} = 100 cover letters from Study 2, and rated them on the 5 principles. The average of these two ratings correlated about as highly with the computer ratings (\textit{r} = .70, \textit{p} < .001) as they did with the average interrater reliability (\textit{r} = .74, \textit{p} < .001, difference in rs, \textit{p} = .56,). These ratings had high internal consistency (Cronbach's $\alpha$s = 0.81, 0.78, and 0.83 for studies 2, 3, and 4, respectively).

To address concerns that particular LLMs might be biased in favor of their own output, we also used Claude to rate the cover letters. We find that GPT ratings correlate strongly with Claude ratings (\textit{r} = .71, \textit{p} < .001), and that the effects are not attenuated by using different models (See Tables \ref{tab:s2_test}, \ref{tab:test_effects_s5}, and \ref{tab:s3_test}), suggesting that our effects are not explainable by same-model bias.

Additionally, we asked a separate sample of participants to read pairs of test-phase cover letters randomly selected from different conditions, and to indicate which letter would be more likely to secure a job interview. Cover letters more likely to secure an interview obtained higher AI ratings of writing skill (\textit{r}s = .29, .35, and .28, \textit{p}s < .001, for Study 2, 3, and 4, respectively).

\subsection*{Statistical analysis}
For Study 1, we conducted one-sample t-tests comparing mean responses to the scale midpoint. 

For Studies 2--4, as per our pre-registrations, we fit ANCOVA models predicting outcomes from condition indicators, controlling for pretest score and baseline characteristics (age, gender, race/ethnicity, primary language, education level, motivation to improve writing skills, self-rated writing skill, experience with AI writing assistants, and baseline writing effectiveness). We used logistic regression to predict whether participants chose to see optional test feedback from condition, controlling for pretest score and baseline characteristics. 
Our analyses of the hypothetical hiring situation use beta regression, because the relative likelihood of a cover letter being preferred is bounded between 0 and 1.
When correcting for multiple comparisons in exploratory moderation analyses, we used the Benjamini-Hochberg correction \cite{benjamini1995controlling}. 

\section*{References}
\bibliography{refs}
\showacknow
\onecolumn
\newpage

\newpage

\appendix
\begin{appendices}
\normalsize

\tableofcontents

\renewcommand{\thefigure}{S\arabic{figure}}
\renewcommand{\thetable}{S\arabic{table}}
\renewcommand{\thesubsection}{\thesection\arabic{subsection}}

\setcounter{figure}{0}
\setcounter{table}{0}
\setcounter{page}{1}
\newpage
\section{Additional methods}

\subsection{AI Ratings}
After participants completed the procedure outlined in Figure \ref{fig:design}, we had three writing samples for participants who practiced with or without AI (one for each phase of the experiment: pretest, practice, test), and two writing samples for participants who did not practice, or simply saw an AI-generated example. We used GPT-4o to rate these texts for five writing principles. Each text and rating was completed independently of each other (i.e., the model had no memory of seeing that text before or of having rated it for any of the other writing principles). For robustness checks, we also used Anthropic's Claude Haiku.

Table \ref{tab:prompts} shows the prompts used to have GPT-4o and Claude rate the rewritten cover letters on the five principles. Our pre-registered main outcome is the unweighted mean of these five principles.

\begin{table}[h]
    \centering
    \caption{Prompt instructions given to GPT-4o and Claude for rating cover letters.}
    \begin{tabular}{lp{15cm}}
    \toprule
    \textbf{Writing principle} & \multicolumn{1}{c}{\textbf{Rating prompt}}\\
    \midrule
        Less is more	& On a 0 $-$ 10 scale, how much does the text follow the Less is more principle? The text should use as few words as needed, as few ideas as needed, and make as few requests as needed. \\
        Easy reading	& On a 0 $-$ 10 scale, how much does the text make reading easy. The text should use short and common words, use straightforward sentences, and shorter sentences.\\
        Easy navigation	& On a 0 $-$ 10 scale, how much does the text make navigation easy. The text should make key information immediately visible, separate distinct ideas, place related ideas together, order ideas by priority, include headings when necessary, and use visuals if needed.\\
        Formatting	& On a 0 $-$ 10 scale, how much does the text use appropriate formatting. The text should follow readers expectations regarding formatting, use bolding, italics, underline, or highlight to draw attention to the most important ideas, and should not overdo formatting.\\
        Easy responding &	On a 0 $-$ 10 scale, how much does the text make responding easy. The text should simplify the steps required to act, organize the key information needed for action, and minimize the amount of attention required.  \\ \bottomrule

    \end{tabular}
    \label{tab:prompts}
\end{table}

\begin{figure}[H]
    \centering
    \includegraphics[width=0.75\linewidth]{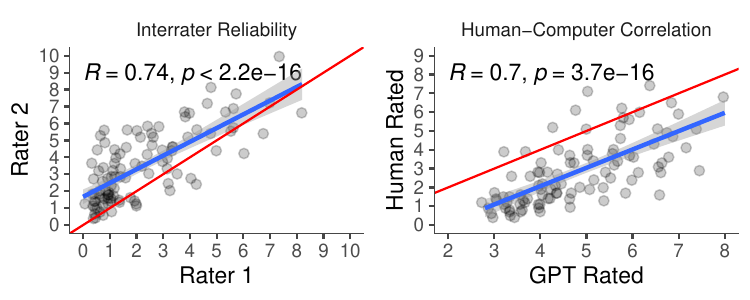}
    \caption{Interrater correlations and correlations between AI and human ratings.}
    \label{fig:ai_valid}
\end{figure}

We used a randomly selected sample of 100 cover letters from Study 2 to validate the AI ratings. Two trained raters independently rated each of the 5 principles. The inter-rater correlation was \textit{r} = .74, \textit{p} < .001. The human raters correlated with the AI-generated ratings satisfactorily (\textit{r} = .70, \textit{p} < .001). The correlation indicating AI-human agreement was not smaller than the interrater agreement correlation (difference in \textit{r}s = .04, \textit{p} = .56), suggesting that AI ratings achieved similar levels of consistency as human raters.   See Figure \ref{fig:ai_valid}.

Claude and GPT-4o ratings were positively correlated, with pretest ratings being less so. In Study 2, the correlations were .38, .78, .65, and .67 for pretest, practice, test, and follow-up, respectively. All \textit{p}-values were below .001.

\subsection{Pairwise Comparisons}
Prolific participants were shown pairs of cover letters sampled from different conditions. 
They were asked to ``Imagine you’re hiring a social media manager for your company; which cover letter would make you more likely to offer an interview to the candidate? Choose one.''
Each cover letter was compared to at least three other letters, sampled uniformly at random from the other two conditions. 
Most letters were compared against 3 or 4 other letters.
For each cover letter, we calculated the relative likelihood of it securing a hypothetical interview, defined as the total number of times that letter was preferred, divided by the total number of contests for that cover letter.

\subsection{Feedback}\label{sec:feedback}

In Studies 2 and 3, participants received feedback for their submissions. The feedback was displayed immediately after the practice cover letter submission. The feedback page read: ``Here is the email, then reproduced the participant's submission verbatim, then read ``Here is one way in which it could be made better.'' The feedback was personalized and created by the GPT-4o model. 

The feedback prompt is shown in Figure \ref{fig:feedback prompt}.
\begin{figure}[H]
    \centering
    \begin{tcolorbox}[colback=blue!5!white, colframe=blue!75!black, title=Feedback prompt]
Take into account the following principles. 
\begin{enumerate}
    \item Less is more (use fewer words, include fewer ideas, make fewer requests).

\item Make reading easy (use short and common words, write straightforward sentences, write shorter sentences).

\item Design for easy navigation (make information immediately visible, group related ideas together, order ideas by priority, include headings).

\item Use enough formatting but no more (match formatting to readers' expectations, highlight, bold, or underline the most important ideas, limit your formatting).

\item Make responding easy (simplify the steps required to act, organize key information needed for action, minimize the amount of attention required).
\end{enumerate}

I will show you a text, and I want you to act as a teacher providing feedback to the email, not the student. To do this, identify the principle that the text would benefit the most from implementing.

Your feedback:

- Should be clear, concise.

- Should reference the text wrote directly, Quote it and offer an alternative
    
- Start with something nice to say about the text
\vspace{1em}

You can structure it as follows:

One sentence about what was good.
	
The email could be improved by focusing on **principle explained concretely in simple words**. For example:

- The email said: **example**

- Instead, it could have said: **rewritten example** 

Make sure the feedback never addresses the person, but always focuses on the text. Never refer to you or your.

One sentence explanation, positive tone.

\end{tcolorbox}
    \caption{Feedback prompt}
    \label{fig:feedback prompt}
\end{figure}

\newpage
\section{Results Study 1}\label{sec:gallup}

We partnered with the Gallup organization to conduct a nationally representative survey of young adults (18-28 years old, N = 2,472). All analyses use survey weights to ensure representativeness of the U.S. population.

\subsection{Usage}
We queried participants about the frequency with which they used AI for different purposes. Table \ref{tab:uses} shows the percentage of participants who engaged in these uses at least once in the past month.

\begin{table}[H]
\caption{Percentage of young adults who used AI for various purposes in the past month}
\centering
\begin{tabular}{lr}
\toprule
\textbf{Use Case} & Percentage\\
\midrule
Google replacement & 64.7\\
Work & 51.5\\
Writing & 45.9\\
Image generation & 39.3\\
Life advice & 32.2\\
Math & 29.8\\
Friend/companionship & 22.9\\
Cheating & 16.4\\
Romantic advice & 9.9\\
\bottomrule
\end{tabular}
\label{tab:uses}
\end{table}

\subsection{Attitudes}

Participants answered four items assessing their beliefs about AI and its effects on human capabilities: “AI makes people less smart”,
“AI reduces people’s ability to learn”,
“AI makes people more smart” (reverse-coded), and
“AI makes people lazier”.

The four items showed good internal consistency (Cronbach’s $\alpha$ = 0.84; weighted $\alpha$ = 0.80).

All four items showed significantly more negative attitudes than the scale midpoint (all \textit{p}s < .001). Specifically, 63\% agreed that AI makes people less smart, 62\% agreed that AI reduces the ability to learn, and 79\% agreed that AI makes people lazier. Overall, 73\% of respondents scored above the midpoint on a composite variable composed of these 4 items, indicating predominantly negative attitudes. See Table \ref{tab:midpoint_comp}

\begin{table}[H]
\centering
\caption{One-sample $t$-tests comparing perceived AI impact to the scale midpoint ($3.5$).}
\begin{tabular}{lcccccc}
\toprule
\textbf{Item} & \textbf{M} & \textbf{SE} & \textbf{$t$} & \textbf{df} & \textbf{$p$} & \textbf{Cohen's $d$} \\
\midrule
``AI makes people less smart''        & 3.93 & 0.04 & 10.3  & 2636 & $< .001$ & 0.30 \\
``AI reduces ability to learn''       & 3.99 & 0.04 & 11.8  & 2636 & $< .001$ & 0.34 \\
``AI makes people more smart'' (rev.) & 2.80 & 0.04 & -18.9 & 2636 & $< .001$ & -0.55 \\
``AI makes people lazier''            & 4.37 & 0.04 & 22.6  & 2636 & $< .001$ & 0.66 \\
\midrule
AI Impact Score (composite)           & 4.12 & 0.03 & 20.4  & 2636 & $< .001$ & 0.57 \\
\bottomrule
\end{tabular}
\label{tab:midpoint_comp}
\end{table}

There was no significant relationship between age and the composite measure of AI Impact Score ($\beta$ = 0.005, p = 0.507).

\section{Results Study 2}

\subsection{Randomization, Balance, and Missingness}
To allow users to format their responses flexibly, we used the TinyMCE rich text editor, which is interfaced with Qualtrics. While this allowed users to use bolding, lists, and italicizing, a small percentage of users experienced technical issues that resulted in their text data not being recorded (3.31\%). These users did type in the box, as evidenced by their time and keystroke data, and completed the experiment. 

There was also attrition in the follow-up sample. While most people responded, 13.45\% of recontacted participants did not respond. This attrition was not selective by condition. As shown in Table \ref{tab:s2missingness}, missingness and attrition rates were low for the main and follow-up samples, and did not differ by condition.

\begin{table}[h]
    \centering
    \caption{Missingness and attrition proportions and test in Study 2.}

\begin{tabular}{lrr}
\toprule
Condition & Main Sample & Follow-up Sample \\ 
\midrule
No practice & $2.25\%$ & $13.38\%$ \\ 
Practice w/o AI & $3.86\%$ & $12.77\%$ \\ 
Practice w/ AI & $3.83\%$ & $14.23\%$ \\ 
\midrule
Overall & $3.31\%$ & $13.45\%$ \\ 
\midrule
$\chi^2$ & $3.966$ & $0.685$ \\ 
\textit{p}-value & $0.138$ & $0.710$ \\ 
\bottomrule
\end{tabular}

    \label{tab:s2missingness}
\end{table}

Pre-treatment variables were balanced across experimental conditions, ensuring that random assignment was successful. To assess balance, we conducted a series of one-way ANOVAs for continuous variables and chi-square tests for categorical variables. Given the multiple comparisons, we applied the Benjamini-Hochberg (BH) procedure to control the false discovery rate. All statistical tests confirmed that none of the pre-treatment variables differed significantly across conditions. See Table \ref{tab:s2randomization}.
    
    \begin{table}[H]
        \centering
            \caption{Randomization checks for pre-treatment variables in Study 2. \textit{p}-values are BH multiple comparisons corrected. Continuous variables tested with ANOVA, binary and factor variables with $\chi^2$ tests. SMD = Standardized Mean Difference.}
        \label{tab:s2randomization}
    \begin{tabular}{lrrrrrrr}
    \toprule
    \multicolumn{1}{l}{} & Overall & No practice & Practice w/o AI & Practice w/ AI & \textit{p} &  SMD \\ 
    \midrule
    \textit{\textbf{n}} & 2238 &  755 &  752 &  731 &  &   \\ 
    \textbf{Age (mean (SD))} & 36.22 (12.71) & 35.99 (13.01) & 36.39 (12.70) & 36.29 (12.42) & .923  &  0.021 \\ 
    \textbf{Gender (\%)} &  &   &   &   & .636 &   0.077 \\ 
    \hspace{1em}   Female & 1189 (53.1)  &  421 (55.8)  &  394 (52.4)  &  374 (51.2)  &  &  \\ 
       \hspace{1em}Male & 1027 (45.9)  &  328 (43.4)  &  348 (46.3)  &  351 (48.0)  &  &    \\ 
       \hspace{1em}Other &   22 (1.0)  &    6 (0.8)  &   10 (1.3)  &    6 (0.8)    &  &  \\ 
       \textbf{Race/Ethnicity} &  &   &   &   &  &  \\
    \hspace{1em}White (\%) & 1288 (57.6)  &  419 (55.5)  &  458 (60.9)  &  411 (56.2)  & .213 &  0.073 \\ 
    \hspace{1em}Black (\%) &  745 (33.3)  &  262 (34.7)  &  223 (29.7)  &  260 (35.6)  & .192 &  0.084 \\ 
    \hspace{1em}Asian (\%) &  134 (6.0)  &   49 (6.5)  &   42 (5.6)  &   43 (5.9)  & .923   &  0.025 \\ 
    \hspace{1em}Latino (\%) &  155 (6.9)  &   48 (6.4)  &   65 (8.6)  &   42 (5.7)  & .213   &  0.075 \\
    \hspace{1em}Other (\%) &   62 (2.8)  &   23 (3.0)  &   22 (2.9)  &   17 (2.3)  & .923 &   0.030 \\ 
    \textbf{Education Level (\%)} &  &   &   &   & .923 &    0.099 \\ 
       \hspace{1em}Less than high school degree &   14 (0.6)  &    5 (0.7)  &    5 (0.7)  &    4 (0.5)  &    &  \\ 
       \hspace{1em}High school graduate (high school diploma or equivalent including GED) &  207 (9.2)  &   68 (9.0)  &   72 (9.6)  &   67 (9.2)  &  &  \\ 
       \hspace{1em}Some college but no degree &  321 (14.3)  &  117 (15.5)  &  109 (14.5)  &   95 (13.0)  &  &   \\ 
       \hspace{1em}Associate degree in college (2-year) &  168 (7.5)  &   61 (8.1)  &   60 (8.0)  &   47 (6.4)    &  &  \\ 
       \hspace{1em}Bachelor's degree in college (4-year) &  984 (44.0)  &  314 (41.6)  &  334 (44.4)  &  336 (46.0)  &  &    \\ 
       \hspace{1em}Master's degree &  474 (21.2)  &  165 (21.9)  &  153 (20.3)  &  156 (21.3)  &  &    \\ 
       \hspace{1em}Doctoral degree (PhD) &   44 (2.0)  &   16 (2.1)  &   11 (1.5)  &   17 (2.3)  &  &    \\ 
       \hspace{1em}Non-PhD Professional degree (JD, MD) &   26 (1.2)  &    9 (1.2)  &    8 (1.1)  &    9 (1.2)  &    &  \\ 
    \textbf{Perceived Writing Skill (mean (SD))} & 6.70 (1.70) & 6.77 (1.67) & 6.56 (1.72) & 6.76 (1.69) & .192 &    0.083 \\ 
    \textbf{Motivation to improve writing (\%)} &  &   &   &   & .410 &    0.126 \\ 
       \hspace{1em}Not at all motivated &   33 (1.5)  &    6 (0.8)  &   15 (2.0)  &   12 (1.6)  &  &    \\ 
       \hspace{1em}Hardly motivated &  106 (4.7)  &   38 (5.0)  &   34 (4.5)  &   34 (4.7)  &  &    \\ 
       \hspace{1em}Somewhat motivated &  644 (28.8)  &  218 (28.9)  &  236 (31.4)  &  190 (26.0)  &  &    \\ 
       \hspace{1em}Very motivated &  932 (41.6)  &  311 (41.2)  &  311 (41.4)  &  310 (42.4)  &  &    \\ 
       \hspace{1em}Extremely motivated &  523 (23.4)  &  182 (24.1)  &  156 (20.7)  &  185 (25.3)  &  &    \\ 
    \textbf{Experience with AI writing assistants (\%)} &  &   &   &   & .701   &  0.094 \\ 
       \hspace{1em}I have never tried any AI writing assistant &  354 (15.8)  &  109 (14.4)  &  139 (18.5)  &  106 (14.5)  &    &  \\ 
       \hspace{1em}I have tried AI writing assistant(s) but hardly ever use them &  859 (38.4)  &  291 (38.5)  &  288 (38.3)  &  280 (38.3)  &  &    \\ 
       \hspace{1em}I use AI writing assistant(s) a few times per week &  477 (21.3)  &  164 (21.7)  &  147 (19.5)  &  166 (22.7)  &  &    \\ 
       \hspace{1em}I use AI writing assistant(s) about once a week &  395 (17.6)  &  136 (18.0)  &  133 (17.7)  &  126 (17.2)  &  &    \\ 
       \hspace{1em}I use AI writing assistant(s) every day &  153 (6.8)  &   55 (7.3)  &   45 (6.0)  &   53 (7.3)  &  &    \\ 
    \textbf{Pretest Writing Skill (mean (SD))} & 3.32 (0.78) & 3.31 (0.73) & 3.32 (0.78) & 3.32 (0.83) & .923 &    0.013 \\ 
    \bottomrule
    \end{tabular}
    \end{table}

\subsection{AI practice improved writing skill}

The AI tool improved performance while participants used it. Table \ref{tab:s2_practice} shows means and standardized differences for different measures of writing skill during the practice phase. The robustness checks included after the main specification, show that results are similar when using a different language model (Column 2), when not including control variables (Column 3), when excluding participants who admitted to cheating in the test phase (Column 4), for the subset of non-attriting participants to the follow-up phase (Column 5), and for each of the 5 principles separately (Columns 6 - 10).

\begin{figure}[H]
    \centering    \includegraphics[width=0.75\linewidth]{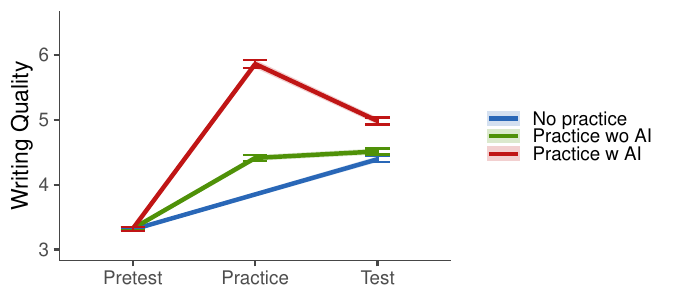}     
    \caption{Participants who had practiced with the AI tool outperformed those who had practiced without it and those who had not practiced at all.
    Error bars represent means $\pm$ 1 SE.
    (\textit{N} = 2,238).}
    \label{fig:s2_supplement}
\end{figure}

\begin{table}[H]
    \centering
    \caption{Differences in writing quality by condition in the practice phase}

\begin{tabular}{lcccccccccc}
\toprule
  & GPT-4o & Claude & Ex. Controls & Ex. Cheaters & Follow-up & LM & ER & EN & F & ER \\ 
\midrule
\multicolumn{11}{l}{\textbf{Means --- (SE)}} \\ 
\midrule
Practice w/o AI & 4.58 & 6.52 & 4.41 & 4.42 & 4.27 & 4.29 & 6.31 & 5.65 & 3.38 & 4.26 \\ 
 & (.222) & (.091) & (.054) & (.055) & (.281) & (.230) & (.150) & (.258) & (.480) & (.268) \\ 
Practice w/ AI & 6.05 & 7.01 & 5.86 & 5.89 & 5.78 & 5.56 & 7.11 & 6.75 & 6.30 & 5.54 \\ 
 & (.224) & (.092) & (.055) & (.055) & (.286) & (.232) & (.151) & (.260) & (.484) & (.271) \\ 
\midrule
\multicolumn{11}{l}{\textbf{Effect Sizes (d) --- (SE)}} \\ 
\midrule
Practice w/o AI vs. Practice w/ AI & 1.01*** & .81*** & .98*** & 1.00*** & 1.05*** & .84*** & .82*** & .65*** & .93*** & .73*** \\ 
 & (.056) & (.055) & (.055) & (.056) & (.061) & (.055) & (.055) & (.054) & (.056) & (.055) \\ 
\midrule
\multicolumn{11}{p{16.5cm}}{\textit{Note.} GPT-4o is the main specification. Ex. Controls is the main specification, unadjusted for demographic and pretreatment variables, Ex. Cheaters excludes the 3\% of participants who admitted to cheating on the test phase. Follow-up is the subsample of non-attriting participants who returned to the one-day follow-up. LM to ER are disaggregated scores for each of the five principles. LM = Less is More, ER = Easy Reading, EN = Easy Navigation, F = Formatting, ER = Easy Responding.  *** \textit{p} < .001, ** \textit{p} < .01, * \textit{p} < .05.}
\vspace{5pt}
\end{tabular}
    \label{tab:s2_practice}
\end{table}

During the test phase, when participants had to rewrite a cover letter without the help of the AI tool, participants who had practiced with AI outperformed participants who had not practiced, or had practiced without the AI tool. Again, the learning gains are robust to different specifications, subsamples, and measures or writing quality. See Table \ref{tab:s2_test}.
For participants assigned to practice with the AI tool, the quality of AI rewrites did not correlate with participants' final submissions, \textit{r} = .06, \textit{p} = .25.

\begin{table}[H]
    \centering
    \caption{Differences in writing quality by condition in the test phase}
\begin{tabular}{lcccccccccc}
\toprule
  & GPT-4o & Claude & Ex. Controls & Ex. Cheaters & Follow-up & LM & ER & EN & F & ER \\ 
\midrule
\multicolumn{11}{l}{\textbf{Means --- (SE)}} \\ 
\midrule
No practice & 4.41 & 6.70 & 4.39 & 4.39 & 4.52 & 3.71 & 5.90 & 5.55 & 2.47 & 4.44 \\ 
 & (.161) & (.072) & (.047) & (.048) & (.200) & (.155) & (.143) & (.192) & (.394) & (.202) \\ 
Practice w/o AI & 4.53 & 6.74 & 4.51 & 4.52 & 4.63 & 3.84 & 6.03 & 5.56 & 2.76 & 4.45 \\ 
 & (.160) & (.071) & (.048) & (.048) & (.199) & (.154) & (.142) & (.190) & (.392) & (.200) \\ 
Practice w/ AI & 5.01 & 6.90 & 4.98 & 4.99 & 5.11 & 4.12 & 6.21 & 6.17 & 3.86 & 4.66 \\ 
 & (.161) & (.072) & (.049) & (.049) & (.202) & (.155) & (.143) & (.192) & (.394) & (.202) \\ 
\midrule
\multicolumn{11}{l}{\textbf{Effect Sizes (d) --- (SE)}} \\ 
\midrule
No practice vs. Practice w/o AI & .09 & .09 & .09 & .10 & .09 & .10* & .11* & .01 & .09 & .01 \\ 
 & (.053) & (.053) & (.052) & (.053) & (.056) & (.053) & (.053) & (.053) & (.053) & (.053) \\ 
No practice vs. Practice w/ AI & .47*** & .36*** & .46*** & .46*** & .48*** & .34*** & .28*** & .42*** & .46*** & .14** \\ 
 & (.054) & (.053) & (.053) & (.054) & (.057) & (.053) & (.053) & (.053) & (.054) & (.053) \\ 
Practice w/o AI vs. Practice w/ AI & .38*** & .28*** & .36*** & .36*** & .39*** & .23*** & .17** & .41*** & .36*** & .13* \\ 
 & (.054) & (.053) & (.053) & (.054) & (.057) & (.054) & (.053) & (.054) & (.054) & (.053) \\ 
\midrule
\multicolumn{11}{p{16.5cm}}{\textit{Note.} GPT-4o is the main specification. Ex. Controls is the main specification, unadjusted for demographic and pretreatment variables, Ex. Cheaters excludes the 3\% of participants who admitted to cheating on the test phase. Follow-up is the subsample of non-attriting participants who returned to the one-day follow-up. LM to ER are disaggregated scores for each of the five principles. LM = Less is More, ER = Easy Reading, EN = Easy Navigation, F = Formatting, ER = Easy Responding.  *** \textit{p} < .001, ** \textit{p} < .01, * \textit{p} < .05.}
\vspace{5pt}
\end{tabular}
    \label{tab:s2_test}
\end{table}

\subsection{AI practice was less effortful}

Table \ref{tab:effort_practice2} shows OLS models predicting practice effort metrics from practice condition. Results show that participants practicing without AI expended more effort, measured subjectively or objectively, through keystrokes or practice time. As pre-registered, time is square-root-transformed, and keystrokes are log-transformed. Differences are slightly smaller when using untransformed variables.

\begin{table}[H]
    \centering
    \caption{Practice effort differences}
\begin{tabular}{lccccc}
\toprule
  & sqrt(Time) & log(Keystrokes) & Subjective Rating (0 - 10) & Time & Keystrokes \\ 
\midrule
\multicolumn{6}{l}{\textbf{Means --- (SE)}} \\ 
\midrule
Practice w/o AI & 2.37 & 4.31 & 6.52 & 6.76 & 430.95 \\ 
 & (.152) & (.322) & (.291) & (.913) & (57.349) \\ 
Practice w/ AI & 2.30 & 3.36 & 5.93 & 6.62 & 383.38 \\ 
 & (.153) & (.325) & (.293) & (.919) & (57.887) \\ 
\midrule
\multicolumn{6}{l}{\textbf{Effect Sizes (d) --- (SE)}} \\ 
\midrule
Practice w/o AI vs. Practice w/ AI & -.07 & -.45*** & -.31*** & -.02 & -.13* \\ 
 & (.053) & (.054) & (.053) & (.053) & (.053) \\ 
\midrule
\multicolumn{6}{l}{\textit{Note.} *** \textit{p} < .001, ** \textit{p} < .01, * \textit{p} < .05.}
\vspace{5pt}

\end{tabular}

    \label{tab:effort_practice2}
\end{table}

Table \ref{tab:effort_test2} shows OLS models predicting test effort metrics from practice condition. Results show some differences: participants who practiced with AI pressed more keys but reported less subjective effort.

\begin{table}[H]
    \centering
\caption{Test effort differences}
\begin{tabular}{lccccc}
\toprule
  & sqrt(Time) & log(Keystrokes) & Subjective Rating (0 - 10) & Time & Keystrokes \\ 
\midrule
\multicolumn{6}{l}{\textbf{Means --- (SE)}} \\ 
\midrule
No practice & 2.32 & 5.01 & 6.55 & 5.62 & 399.10 \\ 
 & (.069) & (.213) & (.266) & (.257) & (41.391) \\ 
Practice w/o AI & 2.15 & 4.87 & 6.91 & 4.98 & 409.09 \\ 
 & (.068) & (.211) & (.265) & (.255) & (41.143) \\ 
Practice w/ AI & 2.19 & 5.05 & 6.69 & 5.14 & 446.59 \\ 
 & (.069) & (.213) & (.267) & (.257) & (41.408) \\ 
\midrule
\multicolumn{6}{l}{\textbf{Effect Sizes (d) --- (SE)}} \\ 
\midrule
No practice vs. Practice w/o AI & -.31*** & -.09 & .18*** & -.32*** & .03 \\ 
 & (.053) & (.053) & (.053) & (.053) & (.053) \\ 
No practice vs. Practice w/ AI & -.24*** & .02 & .07 & -.24*** & .15** \\ 
 & (.053) & (.053) & (.053) & (.053) & (.053) \\ 
Practice w/o AI vs. Practice w/ AI & .07 & .11* & -.11* & .08 & .12* \\ 
 & (.053) & (.053) & (.053) & (.053) & (.053) \\ 
\midrule
\multicolumn{6}{l}{\textit{Note.} *** \textit{p} < .001, ** \textit{p} < .01, * \textit{p} < .05.}
\vspace{5pt}

\end{tabular}
    
    \label{tab:effort_test2}
\end{table}

Table \ref{tab:learning_rate2} shows OLS models predicting learning rate metrics from practice condition. Learning rate is defined as the difference between test and pretest, divided by the effort metric. It shows how many points (10-point scale) the participant improved per unit effort (e.g., per minute spent practicing). Participants who practiced with AI improved their skill more efficiently.

\begin{table}[H]
    \centering
    \caption{Learning rate differences. Means are the rate of improvement per unit sqrt(time (min)), log(keystrokes), subjective rating, raw time in minutes, and raw keystrokes. }

\begin{tabular}{lccccc}
\toprule
  & sqrt(Time) & log(Keystrokes) & Subjective Rating (0 - 10) & Time & Keystrokes \\ 
\midrule
\multicolumn{6}{l}{\textbf{Means --- (SE)}} \\ 
\midrule
Practice w/o AI & .28 & .12 & .27 & .32 & .43 \\ 
 & (.062) & (.094) & (.038) & (.096) & (.073) \\ 
Practice w/ AI & .43 & .36 & .37 & .61 & .62 \\ 
 & (.062) & (.094) & (.038) & (.097) & (.073) \\ 
\midrule
\multicolumn{6}{l}{\textbf{Effect Sizes (d) --- (SE)}} \\ 
\midrule
Practice w/o AI vs. Practice w/ AI & .39*** & .40*** & .41*** & .47*** & .40*** \\ 
 & (.054) & (.054) & (.054) & (.054) & (.054) \\ 
\midrule
\multicolumn{6}{l}{\textit{Note.} *** \textit{p} < .001, ** \textit{p} < .01, * \textit{p} < .05.}
\vspace{5pt}
\end{tabular}
    \label{tab:learning_rate2}
\end{table}

\begin{figure}[H]
    \centering
    \includegraphics[width=0.95\linewidth]{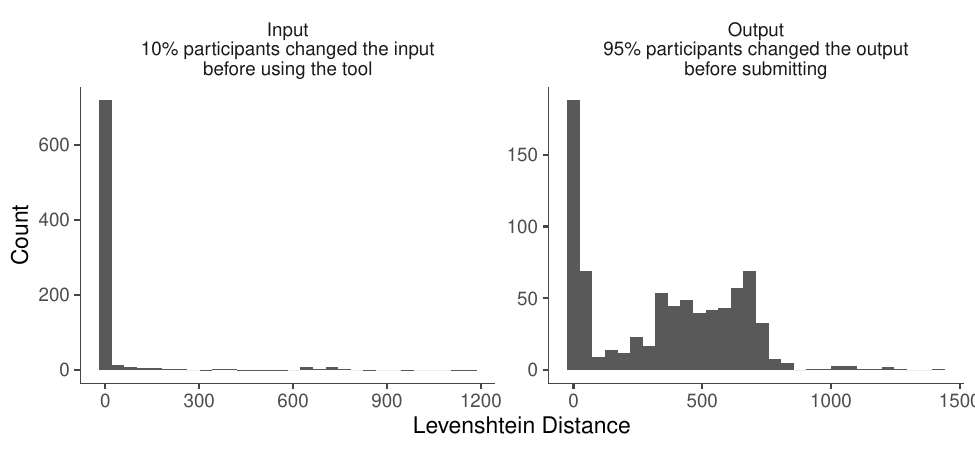}
    \caption{Levenshtein distance (number of additions, modifications or deletions) between the original text and the text passed along to the AI tool (Input); and between the AI's output text and what users submitted as their final work (Output).}
    \label{fig:distance}
\end{figure}

Most participants did not engage passively with the AI tool. As shown in Figure \ref{fig:distance}, an overwhelming majority of participants changed the AI tool's output text before submitting it as their answer. A smaller proportion of participants even edited the cover letter email \textit{before} passing it along to the AI tool.

\subsection{AI practice did not discourage motivation for future learning}\label{sec:future_learning2}

Table \ref{tab:motivation2} presents differences in perceived learning, perceived writing skill, and the likelihood of asking for feedback across conditions, with effect sizes and means reported for each comparison. Despite objectively learning more, participants who practiced with AI perceived their learning and skill levels to be similar to those who practiced without AI and asked for feedback at comparable rates.

\begin{table}[H]
    \centering
    \caption{Differences in motivational variables by condition.}

\begin{tabular}{lccc}
\toprule
  & Perceived learning & Perceived writing skill & Asked for feedback \\ 
\midrule
\multicolumn{4}{l}{\textbf{Means/Proportions}} \\ 
\midrule
No practice & 5.91 & 6.40 & .64 \\ 
 & (.209) & (.208) & (.064) \\ 
Practice w/o AI & 5.90 & 6.61 & .62 \\ 
 & (.207) & (.206) & (.066) \\ 
Practice w/ AI & 6.03 & 6.56 & .58 \\ 
 & (.209) & (.208) & (.068 ) \\ 
\midrule
\multicolumn{4}{l}{\textbf{Effect Sizes (\textit{d}s/odds ratios)}} \\ 
\midrule
No practice vs. Practice w/o AI & -.00 & .13* & 1.10 \\ 
 & (.053) & (.053) & (.128) \\ 
No practice vs. Practice w/ AI & .08 & .10 & 1.30* \\ 
 & (.053) & (.053) & (.149) \\ 
Practice w/o AI vs. Practice w/ AI & .08 & -.03 & 1.17 \\ 
 & (.053) & (.053) & (.134) \\ 
\midrule
\multicolumn{4}{l}{\textit{Note.} *** \textit{p} < .001, ** \textit{p} < .01, * \textit{p} < .05.}
\vspace{5pt}
\end{tabular}
    \label{tab:motivation2}
\end{table}

\subsection{The benefits of practicing with AI were just as large a day later}
\label{sec:persists2}
Table \ref{tab:s2_followup} shows means and standardized differences for measures of writing skill and related outcomes during the follow-up phase. The main specification demonstrates that participants who practiced with AI continued to outperform those who did not practice or practiced without AI. Robustness checks, including using a different language model (Column 2), excluding control variables (Column 3), and removing participants who admitted to cheating (Column 4) confirm the consistency of these effects. The results also hold when evaluating each of the five principles separately (Columns 5–9). These findings suggest that the benefits of practicing with AI are durable and persist even after participants stop using the tool.
\begin{table}[H]
    \centering
    \caption{Differences in writing quality by condition in the follow-up phase}

\begin{tabular}{lccccccccc}
\toprule
  & GPT-4o & Claude & Ex. Controls & Ex. Cheaters & LM & ER & EN & F & ER \\ 
\midrule
\multicolumn{10}{l}{\textbf{Means --- (SE)}} \\ 
\midrule
No practice & 4.73 & 6.75 & 4.75 & 4.75 & 4.31 & 6.36 & 5.56 & 2.43 & 5.01 \\ 
 & (.212) & (.094) & (.054) & (.054) & (.201) & (.206) & (.235) & (.510) & (.267) \\ 
Practice w/o AI & 4.79 & 6.78 & 4.84 & 4.86 & 4.44 & 6.45 & 5.52 & 2.59 & 4.96 \\ 
 & (.211) & (.093) & (.053) & (.054) & (.200) & (.205) & (.234) & (.507) & (.266) \\ 
Practice w/ AI & 5.34 & 6.95 & 5.35 & 5.37 & 4.72 & 6.67 & 6.14 & 3.85 & 5.30 \\ 
 & (.214) & (.094) & (.055) & (.055) & (.203) & (.208) & (.237) & (.515) & (.270) \\ 
\midrule
\multicolumn{10}{l}{\textbf{Effect Sizes (d) --- (SE)}} \\ 
\midrule
No practice vs. Practice w/o AI & .05 & .06 & .07 & .08 & .11 & .08 & -.02 & .05 & -.03 \\ 
 & (.056) & (.054) & (.055) & (.056) & (.056) & (.056) & (.056) & (.056) & (.056) \\ 
No practice vs. Practice w/ AI & .46*** & .33*** & .44*** & .45*** & .32*** & .25*** & .40*** & .45*** & .17** \\ 
 & (.057) & (.055) & (.056) & (.057) & (.056) & (.056) & (.057) & (.057) & (.056) \\ 
Practice w/o AI vs. Practice w/ AI & .41*** & .28*** & .37*** & .37*** & .22*** & .17** & .42*** & .40*** & .20*** \\ 
 & (.057) & (.055) & (.056) & (.056) & (.056) & (.056) & (.057) & (.057) & (.056) \\ 
\midrule
\multicolumn{10}{p{15cm}}{\textit{Note.} GPT-4o is the main specification. Ex. Controls is the main specification, unadjusted for demographic and pretreatment variables, Ex. Cheaters excludes the 3\% of participants who admitted to cheating on the test phase. LM to ER are disaggregated scores for each of the five principles. LM = Less is More, ER = Easy Reading, EN = Easy Navigation, F = Formatting, ER = Easy Responding. *** \textit{p} < .001, ** \textit{p} < .01, * \textit{p} < .05.}
\vspace{5pt}
\end{tabular}
    \label{tab:s2_followup}
\end{table}

As mentioned in the main text the effect of practicing with AI and seeing an AI example did not become attenuated one day later. See Table \ref{tab:followup_attenuation5}.

\begin{table}[H]
\centering
\caption{OLS model interacting condition with phase (Test vs. Follow-up) shows no attenuation of the effect of practicing with AI}
\begin{tabular}{lrrrr}
\toprule
Term & Estimate & SE & t & \textit{p}-value \\ 
\midrule\addlinespace[2.5pt]
Intercept & -11.768 & 3.394 & -3.467 & 0.001 \\ 
Condition: No practice & -0.104 & 0.067 & -1.541 & 0.123 \\ 
Condition: Practice w/ AI & 0.491 & 0.068 & 7.211 & 0.000 \\ 
Condition: No practice $\times$ Follow-up & 0.035 & 0.098 & 0.359 & 0.720 \\ 
Condition: Practice w/ AI $\times$ Follow-up & 0.039 & 0.099 & 0.399 & 0.690 \\ 
Phase: Follow-up & 0.330 & 0.069 & 4.787 & 0.000 \\ 
Pretest score & 0.394 & 0.026 & 15.220 & 0.000 \\ 
Year of birth & 0.007 & 0.002 & 4.284 & 0.000 \\ 
Gender: Male & 0.000 & 0.041 & -0.012 & 0.990 \\ 
Gender: Other & -0.272 & 0.210 & -1.292 & 0.196 \\ 
Race/ethnicity: White & 0.108 & 0.092 & 1.167 & 0.243 \\ 
Race/ethnicity: Black & -0.198 & 0.097 & -2.048 & 0.041 \\ 
Race/ethnicity: Asian & 0.239 & 0.109 & 2.193 & 0.028 \\ 
Race/ethnicity: Latino & 0.070 & 0.095 & 0.736 & 0.462 \\ 
Race/ethnicity: Other & 0.093 & 0.128 & 0.724 & 0.469 \\ 
Education: High school graduate (high school diploma or equivalent including GED) & 0.775 & 0.274 & 2.824 & 0.005 \\ 
Education: Some college but no degree & 0.699 & 0.271 & 2.579 & 0.010 \\ 
Education: Associate degree in college (2-year) & 0.672 & 0.276 & 2.431 & 0.015 \\ 
Education: Bachelor's degree in college (4-year) & 0.682 & 0.269 & 2.536 & 0.011 \\ 
Education: Master's degree & 0.568 & 0.272 & 2.090 & 0.037 \\ 
Education: Doctoral degree (PhD) & 1.122 & 0.308 & 3.643 & 0.000 \\ 
Education: Non-PhD Professional degree (JD, MD) & 0.474 & 0.332 & 1.429 & 0.153 \\ 
Writing skill & 0.014 & 0.013 & 1.045 & 0.296 \\ 
Motivation: Hardly motivated & -0.362 & 0.194 & -1.865 & 0.062 \\ 
Motivation: Somewhat motivated & -0.237 & 0.174 & -1.360 & 0.174 \\ 
Motivation: Very motivated & -0.338 & 0.174 & -1.937 & 0.053 \\ 
Motivation: Extremely motivated & -0.429 & 0.178 & -2.410 & 0.016 \\ 
Experience: I have tried AI writing assistant(s) but hardly ever use them & 0.145 & 0.061 & 2.354 & 0.019 \\ 
Experience: I use AI writing assistant(s) a few times per week & 0.029 & 0.071 & 0.410 & 0.682 \\ 
Experience: I use AI writing assistant(s) about once a week & 0.040 & 0.072 & 0.549 & 0.583 \\ 
Experience: I use AI writing assistant(s) every day & 0.099 & 0.095 & 1.045 & 0.296 \\ 
\bottomrule
\end{tabular}
\label{tab:followup_attenuation2}
\end{table}

\subsection{AI practice was equally effective across subgroups}

To test for moderation effects of pre-treatment demographic variables, we ran separate linear regressions in which writing skill during the test phase was regressed on condition, the pre-treatment moderator of interest, writing skill at baseline, and an interaction term between the moderator $\times$ condition. After correcting the \textit{p-}values for the interaction terms, none were significant at the .05 level, suggesting that practicing with AI was equally effective across groups.

\begin{table}[H]
    \centering
    \caption{BH-corrected \textit{p}-values for interaction terms from models predicting each outcome from condition interacted with pre-treatment variables.}
    \label{tab:interactions2}
        \begin{tabular}{@{\extracolsep{-4pt}}lcccccccccccccc}
    \toprule
    & \multicolumn{2}{c}{\textbf{Test}} & \multicolumn{2}{c}{\textbf{Follow-Up}} & \textbf{Time Practice} & \textbf{Keys Practice} & \textbf{Effort Practice} & \multicolumn{2}{c}{\textbf{Per. Learning}} & \multicolumn{2}{c}{\textbf{Per. Skill}} & \multicolumn{2}{c}{\textbf{Want Feedback}} \\ 
    \cmidrule(lr){2-3} \cmidrule(lr){4-5} \cmidrule(lr){6-6} \cmidrule(lr){7-7} \cmidrule(lr){8-8} \cmidrule(lr){9-10} \cmidrule(lr){11-12} \cmidrule(lr){13-14}
    \textbf{Level} & \textbf{No AI} & \textbf{With AI} & \textbf{No AI} & \textbf{With AI} & \textbf{With AI} & \textbf{With AI} & \textbf{With AI} & \textbf{No AI} & \textbf{With AI} & \textbf{No AI} & \textbf{With AI} & \textbf{No AI} & \textbf{With AI} \\ 
    \midrule
    
    \multicolumn{14}{l}{\textbf{Continuous Moderators}} \\ 
    \midrule
    Pretest         & 0.955 & 0.914 & 0.631 & 0.699 & 0.984 & 0.914 & 0.941 & 0.820 & 0.914 & 0.914 & 0.914 & 0.574 & 0.851 \\ 
    Year of Birth   & 0.533 & 0.931 & 0.868 & 0.955 & 0.914 & 0.955 & 0.914 & 0.914 & 0.955 & 0.955 & 0.914 & 0.914 & 0.618 \\ 
    Writing Skill   & 0.914 & 0.914 & 0.955 & 0.545 & 0.913 & 0.838 & 0.914 & 0.851 & 0.914 & 0.914 & 0.919 & 0.914 & 0.914 \\ 
    \midrule
    
    \multicolumn{14}{l}{\textbf{Gender} (vs. Female)} \\ 
    \midrule
    Male            & 0.955 & 0.914 & 0.970 & 0.955 & 0.699 & 0.914 & 0.955 & 0.699 & 0.979 & 0.914 & 0.851 & 0.699 & 0.533 \\ 
    Other           & 0.719 & 0.919 & 0.914 & 0.955 & 0.931 & 0.955 & 0.931 & 0.914 & 0.914 & 0.955 & 0.931 & 0.876 & 0.913 \\ 
    \midrule
    
    \multicolumn{14}{l}{\textbf{Race}} \\ 
    \midrule
    White           & 0.914 & 0.955 & 0.914 & 0.574 & 0.955 & 0.914 & 0.914 & 0.737 & 0.931 & 0.643 & 0.851 & 0.533 & 0.699 \\ 
    Black           & 0.914 & 0.913 & 0.914 & 0.851 & 0.914 & 0.914 & 0.955 & 0.574 & 0.919 & 0.295 & 0.699 & 0.574 & 0.533 \\ 
    Asian           & 0.973 & 0.973 & 0.955 & 0.737 & 0.964 & 0.931 & 0.955 & 0.931 & 0.890 & 0.919 & 0.955 & 0.914 & 0.699 \\ 
    Latino          & 0.919 & 0.914 & 0.970 & 0.914 & 0.914 & 0.868 & 0.533 & 0.868 & 0.944 & 0.533 & 0.914 & 0.970 & 0.574 \\ 
    Other           & 0.973 & 0.914 & 0.914 & 0.973 & 0.914 & 0.931 & 0.643 & 0.914 & 0.914 & 0.955 & 0.914 & 0.533 & 0.533 \\ 
    \midrule
    
    \multicolumn{14}{l}{\textbf{Motivation} (vs. Not at all)} \\ 
    \midrule
    Hardly          & 0.914 & 0.914 & 0.955 & 0.955 & 0.914 & 0.955 & 0.964 & 0.914 & 0.914 & 0.955 & 0.931 & 0.699 & 0.533 \\ 
    Somewhat        & 0.876 & 0.663 & 0.973 & 0.913 & 0.914 & 0.914 & 0.914 & 0.973 & 0.964 & 0.914 & 0.955 & 0.663 & 0.533 \\ 
    Very            & 0.851 & 0.566 & 0.980 & 0.914 & 0.851 & 0.968 & 0.964 & 0.970 & 0.982 & 0.914 & 0.973 & 0.699 & 0.533 \\ 
    Extremely       & 0.861 & 0.533 & 0.973 & 0.868 & 0.914 & 0.946 & 0.914 & 0.931 & 0.964 & 0.914 & 0.964 & 0.749 & 0.533 \\ \midrule
    \multicolumn{14}{l}{\textbf{Experience with AI writing assistants} (vs. None)} \\ 
\midrule
Hardly ever & 0.931 & 0.914 & 0.931 & 0.868 & 0.914 & 0.533 & 0.914 & 0.574 & 0.931 & 0.931 & 0.914 & 0.955 & 0.964 \\ 
A few times per week     & 0.914 & 0.667 & 0.955 & 0.955 & 0.914 & 0.533 & 0.914 & 0.574 & 0.919 & 0.931 & 0.914 & 0.914 & 0.955 \\ 
About once a week        & 0.876 & 0.214 & 0.914 & 0.955 & 0.955 & 0.699 & 0.914 & 0.574 & 0.914 & 0.931 & 0.919 & 0.914 & 0.970 \\ 
Every day                & 0.914 & 0.914 & 0.914 & 0.964 & 0.919 & 0.914 & 0.984 & 0.914 & 0.931 & 0.955 & 0.914 & 0.955 & 0.931 \\ 
\midrule
\multicolumn{14}{p{\textwidth}}{\textit{Note.} Models for test and follow-up performance, square-root practice time, log keystrokes, subjective effort, perceived learning and perceived writing skill or OLS models. Asking to see feedback was a binary Yes/No variable, and was modelled with logistic regression. Models match the pre-registered main specification, thereby controlling for all other pre-treatment variables. Per. = Perceived}
\vspace{5pt}
    \end{tabular}
\end{table}

\newpage
\subsection{Pairwise comparisons}
\begin{figure}[H]
    \centering
    \includegraphics[width=0.5\linewidth]{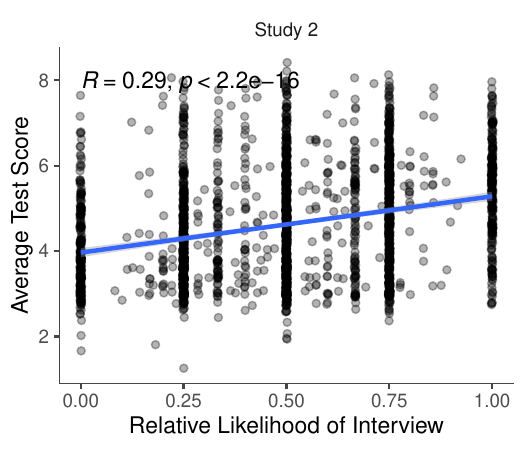}
    \caption{Correlation between test score writing quality as rated by GPT-4o and relative likelihood of being offered a hypothetical interview.}
    \label{fig:correlations2}
\end{figure}

The relative likelihood of a cover letter receiving an invitation to an interview was correlated with the GPT-rated writing quality. See Figure \ref{fig:correlations2}.

As shown in Table \ref{tab:beta2}, participants who had practiced writing cover letters with AI were more likely to be invited to a hypothetical job interview.

\begin{table}[H]
\centering
\caption{Beta regression models predicting relative likelihood of interview from condition.}

\begin{tabular}{lcc}
\toprule
 & Without controls & With controls \\ 
\midrule\addlinespace[2.5pt]
\multicolumn{3}{l}{\textbf{Means (SE)}} \\[2.5pt] 
\midrule\addlinespace[2.5pt]
Practice w AI & .50 & .53 \\ 
 & (.01) & (.01) \\ 
Practice wo AI & .48 & .48 \\ 
 & (.01) & (.01) \\ 
See AI example & .51 & .49 \\ 
 & (.01) & (.01) \\ 
\midrule\addlinespace[2.5pt]
\multicolumn{3}{l}{\textbf{Model coefficients (SE)}} \\[2.5pt] 
\midrule\addlinespace[2.5pt]
Precision ($\phi$) & 16.17*** & 5.36*** \\ 
 & (1.47) & (.30) \\ 
Symmetry (Log($\nu$)) & .35*** & -.91*** \\ 
 & (.08) & (.07) \\ 
\midrule\addlinespace[2.5pt]
\multicolumn{3}{l}{\textbf{Pairwise comparison (SE)}} \\[2.5pt] 
\midrule\addlinespace[2.5pt]
Practice w AI - Practice wo AI & .08** & .21*** \\ 
 & (.03) & (.04) \\ 
Practice w AI - See AI example & -.02 & .14** \\ 
 & (.03) & (.04) \\ 
Practice wo AI - See AI example & -.10*** & -.06 \\ 
 & (.03) & (.04) \\ 
\midrule\addlinespace[2.5pt]
\multicolumn{3}{l}{\textbf{Statistics}} \\[2.5pt] 
\midrule\addlinespace[2.5pt]
N & 1934.00 & 2153.00 \\ 
AIC & 2447.07 & 2329.29 \\ 
BIC & 2474.91 & 2499.52 \\ 
log(Likelihood) & -1218.54 & -1134.64 \\ 
\bottomrule
\end{tabular}
\label{tab:beta2}
\end{table}

\subsection{Forecasting}
Table \ref{tab:ORs1} reports results from a logistic regression predicting the probability of answering correctly as a function of demographic and background characteristics. Odds ratios (ORs) are displayed along with their standard errors, Wald z statistics, and \textit{p}-values. The model includes age, sex, ethnicity, student status, employment, and prior AI experience as predictors.

\begin{table}[H]
    \centering
    \caption{Demographic covariates and odds ratios for correct responses}
\begin{tabular}{@{\extracolsep{\fill}}lrrrr}
\toprule
\textbf{Predictor} & \textbf{OR} & \textbf{SE} & \textbf{z} & \textbf{p-value} \\ 
\midrule\addlinespace[2.5pt]
Prior AI experience & 0.718 & 0.199 & -1.664 & 0.096 \\ 
Age (years) & 1.014 & 0.021 & 0.675 & 0.500 \\ 
Male (vs. Female) & 1.035 & 0.467 & 0.073 & 0.942 \\ 
White (vs. Other) & 0.966 & 0.509 & -0.068 & 0.946 \\ 
Student (vs. Not) & 1.120 & 0.618 & 0.183 & 0.855 \\ 
\textbf{Employment status} & & & & \\
\hspace{1em}Other & 0.770 & 0.687 & -0.382 & 0.703 \\ 
\hspace{1em}Part-time & 0.663 & 0.601 & -0.684 & 0.494 \\ 
\hspace{1em}Unemployed (seeking) & 0.645 & 0.593 & -0.738 & 0.461 \\ 
\bottomrule
\end{tabular}
\label{tab:ORs1}
\end{table}

\newpage

\begin{figure}[]
    \centering
    \includegraphics[width=.5\linewidth]{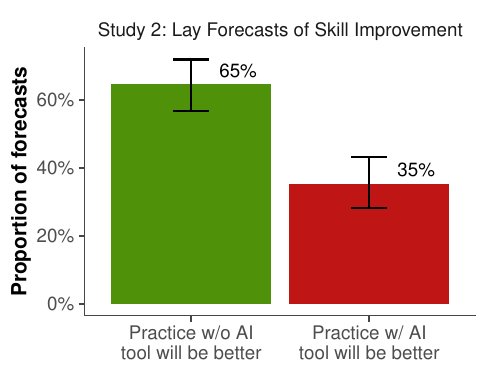}
    \caption{Forecasters in Study 2 predicted that practicing without an AI tool would improve writing skill more than practicing with an AI tool. Error bars represent proportions $\pm$ 1 SE.}
    \label{fig:s1}
\end{figure}

\section{Results Study 3}

\subsection{Randomization, Balance, and Missingness}
Participants in Study 3 were recontacted to return two days after the initial intake section. While most people responded, 14\% of recontacted participants did not respond. As shown in Table \ref{tab:s5missingness}, missingness and attrition rates were low and did not differ by condition.

\begin{table}[H]
\centering
\caption{Missingness and attrition proportions and test in Study 3.}
\begin{tabular}{lr}
\toprule
Condition & Retention Rate \\ 
\midrule
Practice w/ AI & 85.11\% \\ 
Google examples & 85.70\% \\ 
Editor feedback & 86.95\% \\ \midrule
Overall & 85.92\% \\ \midrule
$\chi^2$ & 1.446 \\ 
\textit{p}-value & .485 \\ 
\bottomrule
\end{tabular}
\label{tab:s5missingness}
\end{table}

Pre-treatment variables were balanced across experimental conditions, ensuring that random assignment was successful. To assess balance, we conducted a series of one-way ANOVAs for continuous variables and chi-square tests for categorical variables. Given the multiple comparisons, we applied the Benjamini-Hochberg (BH) procedure to control the false discovery rate. All statistical tests confirmed that none of the pre-treatment variables differed significantly across conditions. See Table \ref{tab:s5randomization}.
    
\begin{table}[H]
\small
            \caption{Randomization checks for pre-treatment variables in Study 3. \textit{p}-values are BH multiple comparisons corrected. Continuous variables tested with ANOVA, binary and factor variables with $\chi^2$ tests. SMD = Standardized Mean Difference.}

\begin{tabular}{lrrrrrrr}
\toprule
 & Overall & Practice w/ AI & Google examples & Editor feedback & \textit{p} &  SMD \\ 
\midrule
\textit{\textbf{n}} &  2997 &  1001 &  1000 &   996 &  &    \\ 
\textbf{Age (mean (SD))} & 41.47 (13.83) & 41.38 (13.78) & 41.11 (13.76) & 41.92 (13.95) & .666 &    0.039 \\ 
\textbf{Gender (\%)} &   &    &    &    & .666 &    0.059 \\ 
\hspace{1em}Female &  1824 (60.9)  &   596 (59.5)  &   628 (62.8)  &   600 (60.2)  &  &    \\ 
\hspace{1em}Male &  1124 (37.5)  &   390 (39.0)  &   352 (35.2)  &   382 (38.4)  &  &   \\ 
\hspace{1em}Other &    49 ( 1.6)  &    15 ( 1.5)  &    20 ( 2.0)  &    14 ( 1.4)  &  &    \\ 
   \textbf{Race/Ethnicity} & & &&&\\
\hspace{1em}White (\%) &  2209 (73.7)  &   745 (74.4)  &   736 (73.6)  &   728 (73.1)  & .950 &    0.020 \\ 
\hspace{1em}Black (\%) &   473 (15.8)  &   164 (16.4)  &   145 (14.5)  &   164 (16.5)  & .666 &    0.036 \\ 
\hspace{1em}Asian (\%) &   204 ( 6.8)  &    61 ( 6.1)  &    68 ( 6.8)  &    75 ( 7.5)  & .666 &   0.038 \\ 
\hspace{1em}Latino (\%) &   253 ( 8.4)  &    79 ( 7.9)  &    95 ( 9.5)  &    79 ( 7.9)  & .666 &   0.038 \\ 
\hspace{1em}Other  (\%) &   102 ( 3.4)  &    35 ( 3.5)  &    40 ( 4.0)  &    27 ( 2.7)  & .666 &   0.048 \\ 
\textbf{Education Level (\%)} &   &    &    &    & .964 &    0.080 \\ 
   \hspace{1em}Less than high school degree &    10 ( 0.3)  &     4 ( 0.4)  &     2 ( 0.2)  &     4 ( 0.4)  &  &    \\ 
   \hspace{1em}High school graduate (high school diploma or equivalent including GED) &   271 ( 9.0)  &    80 ( 8.0)  &    94 ( 9.4)  &    97 ( 9.7)  &  &    \\ 
   \hspace{1em}Some college but no degree &   508 (17.0)  &   178 (17.8)  &   168 (16.8)  &   162 (16.3)  &  &    \\ 
   \hspace{1em}Associate degree in college (2-year) &   283 ( 9.4)  &    91 ( 9.1)  &    98 ( 9.8)  &    94 ( 9.4)  &  &    \\ 
   \hspace{1em}Bachelor's degree in college (4-year) &  1165 (38.9)  &   398 (39.8)  &   394 (39.4)  &   373 (37.4)  &  &    \\ 
   \hspace{1em}Master's degree &   596 (19.9)  &   199 (19.9)  &   193 (19.3)  &   204 (20.5)  &  &    \\ 
   \hspace{1em}Doctoral/Professional degree &   164 ( 5.5)  &    51 ( 5.1)  &    51 ( 5.1)  &    62 ( 6.2)    &  &  \\ 
\textbf{Perceived Writing Skill (mean (SD))} &  6.69 (1.67) &  6.67 (1.68) &  6.63 (1.69) &  6.76 (1.63) & .666 &    0.051 \\ 
\textbf{Motivation to Improve Writing} (\%) &   &    &    &    & .748 &    0.083 \\ 
   \hspace{1em}Not at all motivated &    24 ( 0.8)  &     9 ( 0.9)  &     6 ( 0.6)  &     9 ( 0.9)  &  &    \\ 
   \hspace{1em}Hardly motivated &   166 ( 5.5)  &    59 ( 5.9)  &    55 ( 5.5)  &    52 ( 5.2)  &  &    \\ 
   \hspace{1em}Somewhat motivated &  1036 (34.6)  &   354 (35.4)  &   330 (33.0)  &   352 (35.3)  &  &    \\ 
   \hspace{1em}Very motivated &  1135 (37.9)  &   387 (38.7)  &   390 (39.0)  &   358 (35.9)  &  &    \\ 
   \hspace{1em}Extremely motivated &   636 (21.2)  &   192 (19.2)  &   219 (21.9)  &   225 (22.6)  &  &    \\ 
\textbf{Experience with AI writing assistants (\%)} &   &    &    &    & .666 &   0.090 \\ 
   \hspace{1em}I have never tried any AI writing assistant &   516 (17.2)  &   188 (18.8)  &   173 (17.3)  &   155 (15.6)  &    &  \\ 
   \hspace{1em}I have tried AI writing assistant(s) but hardly ever use them &  1211 (40.4)  &   383 (38.3)  &   411 (41.1)  &   417 (41.9)  &  &    \\ 
   \hspace{1em}I use AI writing assistant(s) a few times per week &   557 (18.6)  &   197 (19.7)  &   173 (17.3)  &   187 (18.8)  &  &    \\ 
   \hspace{1em}I use AI writing assistant(s) about once a week &   503 (16.8)  &   158 (15.8)  &   178 (17.8)  &   167 (16.8)  &  &    \\ 
   \hspace{1em}I use AI writing assistant(s) every day &   210 ( 7.0)  &    75 ( 7.5)  &    65 ( 6.5)  &    70 ( 7.0)  &  &  &  \\ 
\textbf{Pretest Writing Skill (mean (SD))} &  4.40 (0.67) &  4.41 (0.67) &  4.40 (0.68) &  4.40 (0.65) & .964 &    0.008 \\ 
\bottomrule
\end{tabular}
\label{tab:s5randomization}
\end{table}

\subsection{Details on professional editors}\label{sec:editors}
We recruited 44 professional writers and editors through Journalist’s Resource, paying them \$100 to edit 20 cover letters each. They were asked to spend about 7 minutes per cover letter. In total, they edited 1,227 cover letters.

To determine if editors changed their behavior over time, we fit a series of multilevel models, with cover letter position as a predictor, random intercept for editors, and text quality (as rated by GPT-4o), time spent per cover letter, and number of insertions, deletions, or modifications (i.e., Levenshtein distance). These analyses include editors who were able to edit at least 20 letters, and focus on the first 20 letters edited. The quality of the edited letters improved over time ($\beta$ = 0.013, SE = 0.005, p = 0.013), even though editors spent less time as the task progressed ($\beta$ = -0.387, SE = 0.028, \textit{p} < .001). Interestingly, editors also were able to improve the emails using fewer edits as the task progressed ($\beta$ = -2.572, SE = 0.926, \textit{p} = 0.00561). Despite spending less time per letter, editors produced higher quality edits with fewer but more effective changes, demonstrating clear learning effects in professional editing tasks.

\begin{figure}[H]
    \centering
    \includegraphics[width=0.85\linewidth]{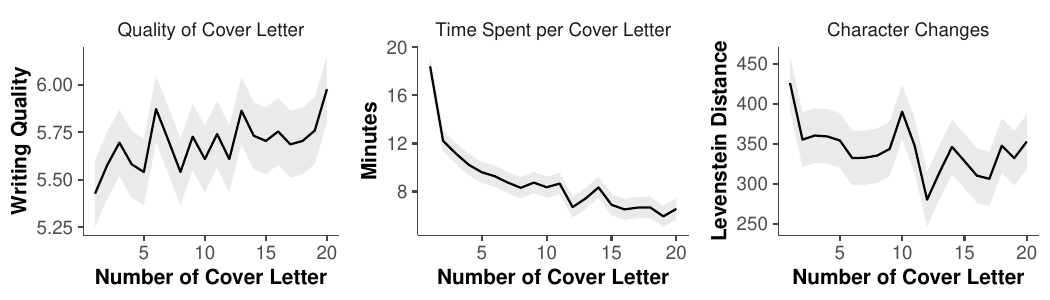}
\caption{Editor performance improved over the course of the task, with participants becoming more efficient at editing emails while reducing both editing time and the number of character-level changes (measured by Levenshtein distance). The line and shaded area show a linear model fit, while the gray line represents the best-fit loess curve.}
    \label{fig:editors_over_time}
\end{figure}

\begin{figure}[H]
    \centering

    \begin{subfigure}{0.48\linewidth}
        \centering
        \includegraphics[width=\linewidth]{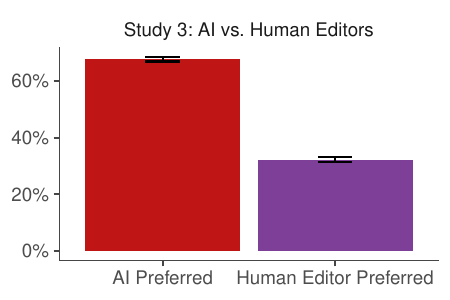}
        \caption{Hypothetical job interviews.}
        \label{fig:winrate}
    \end{subfigure}
    \hfill
    \begin{subfigure}{0.48\linewidth}
        \centering
        \includegraphics[width=\linewidth]{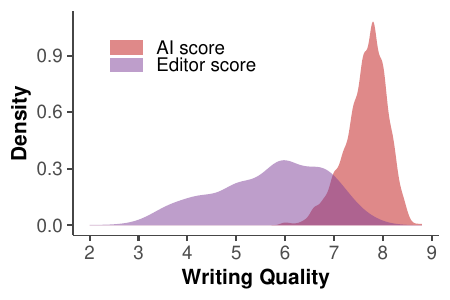}
        \caption{Writing Quality.}
        \label{fig:density}
    \end{subfigure}

    \caption{Counterfactual analysis. For each letter originally assigned to a professional editor, we generated an AI-rewritten counterfactual version, allowing within-letter comparison of AI vs. human editorial quality. Letters rewritten by AI were of highwe quality than those rewritten by professional human editors, as measured by relative likelihood of interview (a) and AI ratings (b).}
    \label{fig:ai_vs_editors}
\end{figure}

Why did participants learn more from AI than they did from professional human editors. One possibility is that the quality of the edits provided by AI were of higher quality than those of human editors. To test this possibility, we used our AI tool to rewrite the same letters that were assigned to professional human editors. That way, for each participant in the ``practice with editor feedback condition'' we have the actual letter they received (as edited by the human expert), and the counterfactual letter they would have received had they been assigned to practice with AI. As shown in Figure \ref{fig:winrate}, a sample of naive human evaluators were twice as likely to prefer hte cover letter edited by AI than the one participants received, edited by a professional human editor (paired \textit{d} = 0.62, \textit{p} < .001) Figure \ref{fig:density} shows the densities of writing quality as evaluated by our AI rating procedure. Cover letters edited by AI were rated as significantly higher quality than those submitted by professional human editors (paired \textit{d} = 1.66, \textit{p} < .001)

\subsection{Details on participants' search behavior}

We replicated the first five pages of a Google search for “cover letter examples” and made this available to participants through Qualtrics to monitor their search behavior.

Across 635 participants assigned to use Google to search for cover letter examples and tips, we observed search behavior consistent with prior research on web browsing \cite{pan2007google}. Most participants (97\%) stayed on page 1, while only 3\% ventured beyond the first page of results. Participants spent an average of 2.8 minutes (median: 1.3 minutes) browsing search results, with an average of 0.9 minutes between clicks. On average, participants made 2.9 clicks during their search session.

\begin{figure}[H]
    \centering
    \includegraphics[width=\linewidth]{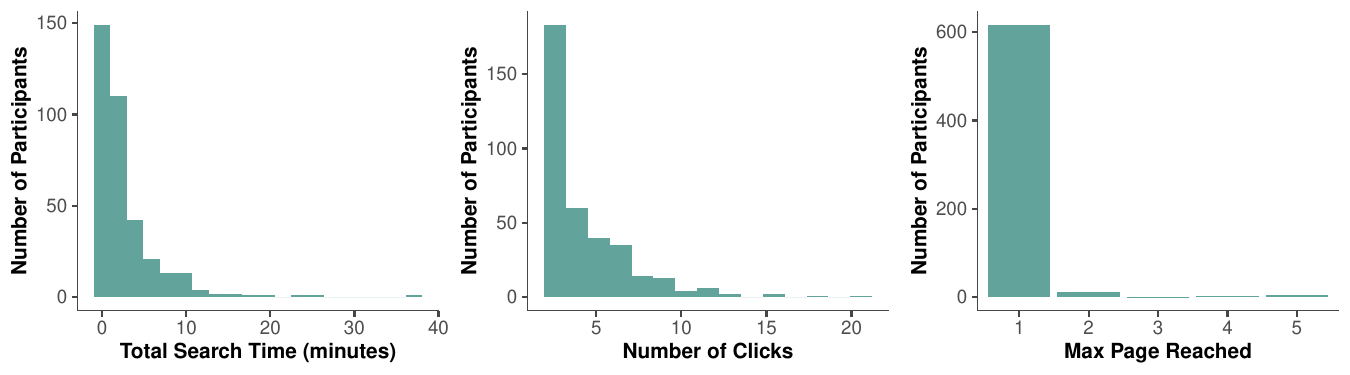}
    \caption{Web browsing descriptive statistics.}
    \label{fig:placeholder}
\end{figure}

\subsection{AI practice improved writing skill more than getting feedback from professional editors and looking for examples online}

The AI tool improved performance while participants used it. Table \ref{tab:practice_effects_s5} shows means and standardized differences for different measures of writing skill during the practice phase. The robustness checks included after the main specification, show that results are similar when using a different language model (Column 2), when not including control variables (Column 3), when excluding participants who admitted to cheating in the test phase (Column 4), and for each of the 5 principles separately (Columns 5 - 9).

\begin{table}[H]
\centering
\caption{Differences in writing quality by condition in the practice phase}
\begin{tabular}{lcccccccc}
\toprule
  & GPT-4o & Ex. Controls & Ex. Cheaters & LM & ER & EN & F & ER \\ 
\midrule
\multicolumn{9}{l}{\textbf{Means}} \\[2.5pt] 
\midrule
{AI} & {6.24} & {6.22} & {6.28} & {5.61} & {7.26} & {6.88} & {5.41} & {6.04} \\ 
{} & {(.172)} & {(.048)} & {(.173)} & {(.167)} & {(.125)} & {(.190)} & {(.397)} & {(.214)} \\ 
{Google} & {4.81} & {4.80} & {4.84} & {4.23} & {6.33} & {5.85} & {2.71} & {4.91} \\ 
{} & {(.173)} & {(.048)} & {(.174)} & {(.167)} & {(.125)} & {(.191)} & {(.398)} & {(.215)} \\ 
{Editors} & {5.18} & {5.17} & {5.21} & {4.55} & {6.56} & {6.24} & {3.44} & {5.09} \\ 
{} & {(.172)} & {(.048)} & {(.173)} & {(.166)} & {(.124)} & {(.190)} & {(.396)} & {(.214)} \\ 
\midrule
\multicolumn{9}{l}{\textbf{Effect Sizes (d)}} \\[2.5pt] 
\midrule
{AI vs. Google} & {-1.04***} & {-1.01***} & {-1.03***} & {-1.03***} & {-.93***} & {-.67***} & {-.85***} & {-.66***} \\ 
{} & {(.051)} & {(.050)} & {(.052)} & {(.051)} & {(.051)} & {(.050)} & {(.051)} & {(.050)} \\ 
{AI vs. Editors} & {-.77***} & {-.75***} & {-.77***} & {-.79***} & {-.70***} & {-.42***} & {-.62***} & {-.56***} \\ 
{} & {(.050)} & {(.049)} & {(.050)} & {(.050)} & {(.050)} & {(.049)} & {(.050)} & {(.050)} \\ 
{Google vs. Editors} & {.27***} & {.26***} & {.27***} & {.24***} & {.23***} & {.25***} & {.23***} & {.10*} \\ 
{} & {(.049)} & {(.048)} & {(.049)} & {(.049)} & {(.049)} & {(.049)} & {(.049)} & {(.049)} \\ 
\midrule
\multicolumn{9}{p{14cm}}{\textit{Note.} GPT-4o is the main specification. Ex. Controls is the main specification, unadjusted for demographic and pretreatment variables, Ex. Cheaters excludes the 3\% of participants who admitted to cheating on the test phase.  LM to ER are disaggregated scores for each of the five principles. LM = Less is More, ER = Easy Reading, EN = Easy Navigation, F = Formatting, ER = Easy Responding.  *** \textit{p} < .001, ** \textit{p} < .01, * \textit{p} < .05.}
\end{tabular}
\label{tab:practice_effects_s5}
\end{table}

During the test phase, when participants had to rewrite a cover letter without the help of the AI tool, participants who had practiced with AI outperformed participants who searched for cover letter examples and tips and participants who received personalized feedback from professional editors. Again, the learning gains are robust to different specifications, subsamples, and measures or writing quality. See Table     \ref{tab:test_effects_s5}.

\begin{table}[H]

    \centering
        \caption{Differences in writing quality by condition in the test phase}
    \begin{tabular}{lccccccccc}
\toprule
  & GPT-4o & Claude & Ex. Controls & Ex. Cheaters & LM & ER & EN & F & ER \\ 
\midrule
\multicolumn{10}{l}{\textbf{Means}} \\[2.5pt] 
\midrule
{AI} & {4.98} & {5.14} & {4.77} & {4.99} & {3.98} & {6.34} & {6.00} & {3.63} & {4.93} \\ 
{} & {(.133)} & {(.103)} & {(.041)} & {(.134)} & {(.126)} & {(.133)} & {(.156)} & {(.333)} & {(.169)} \\ 
{Google} & {4.44} & {4.90} & {4.24} & {4.46} & {3.59} & {5.95} & {5.50} & {2.43} & {4.74} \\ 
{} & {(.133)} & {(.103)} & {(.040)} & {(.134)} & {(.126)} & {(.133)} & {(.156)} & {(.333)} & {(.169)} \\ 
{Editors} & {4.75} & {5.03} & {4.55} & {4.76} & {3.70} & {6.10} & {5.95} & {3.18} & {4.80} \\ 
{} & {(.133)} & {(.103)} & {(.040)} & {(.134)} & {(.126)} & {(.133)} & {(.155)} & {(.333)} & {(.169)} \\ 
\midrule
\multicolumn{10}{l}{\textbf{Effect Sizes (d)}} \\[2.5pt] 
\midrule
{AI vs. Google} & {-.46***} & {-.27***} & {-.44***} & {-.46***} & {-.36***} & {-.34***} & {-.37***} & {-.41***} & {-.12*} \\ 
{} & {(.050)} & {(.050)} & {(.049)} & {(.050)} & {(.049)} & {(.049)} & {(.049)} & {(.049)} & {(.049)} \\ 
{AI vs. Editors} & {-.20***} & {-.12*} & {-.18***} & {-.20***} & {-.25***} & {-.21***} & {-.04} & {-.16**} & {-.09} \\ 
{} & {(.049)} & {(.049)} & {(.048)} & {(.049)} & {(.049)} & {(.049)} & {(.049)} & {(.049)} & {(.049)} \\ 
{Google vs. Editors} & {.26***} & {.15**} & {.26***} & {.26***} & {.11*} & {.13**} & {.34***} & {.26***} & {.04} \\ 
{} & {(.049)} & {(.049)} & {(.048)} & {(.049)} & {(.049)} & {(.049)} & {(.049)} & {(.049)} & {(.049)} \\ 
\midrule
\multicolumn{10}{p{14cm}}{\textit{Note.} GPT-4o is the main specification. Ex. Controls is the main specification, unadjusted for demographic and pretreatment variables, Ex. Cheaters excludes the 3\% of participants who admitted to cheating on the test phase. LM to ER are disaggregated scores for each of the five principles. LM = Less is More, ER = Easy Reading, EN = Easy Navigation, F = Formatting, ER = Easy Responding.  *** \textit{p} < .001, ** \textit{p} < .01, * \textit{p} < .05.}
    \end{tabular}
    \label{tab:test_effects_s5}
\end{table}

\newpage
\subsection{AI practice wasn’t any more effortful than getting feedback from professional editors and was less effortful than looking for examples online}

Table \ref{tab:effort_practice5} shows OLS models predicting practice effort metrics from practice condition. Results show that participants practicing without AI expended more effort, measured subjectively or objectively, through keystrokes or practice time. As pre-registered, time is square-root-transformed, and keystrokes are log-transformed. Differences are slightly smaller when using untransformed variables.

\begin{table}[H]
    \centering
    \caption{Practice effort differences}

    \begin{tabular}{lcccccc}
\toprule
  & sqrt(Time) & log(Keystrokes) & Subjective Rating (0 - 10) & Time & Keystrokes & Keystrokes/min \\ 
\midrule
\multicolumn{7}{l}{Means} \\[2.5pt] 
\midrule
AI & 2.45 & 3.76 & 7.14 & 7.04 & 251.19 & 1.52 \\ 
 & (.112) & (.251) & (.217) & (.702) & (41.527) & (.089) \\ 
Google & 2.76 & 4.68 & 7.11 & 8.96 & 369.45 & 1.78 \\ 
 & (.112) & (.251) & (.217) & (.702) & (41.464) & (.089) \\ 
Editors & 2.54 & 4.50 & 7.27 & 7.50 & 281.33 & 1.83 \\ 
 & (.112) & (.251) & (.216) & (.702) & (41.482) & (.089) \\ 
\midrule
\multicolumn{7}{l}{Effect Sizes (d)} \\[2.5pt] 
\midrule
AI vs. Google & .32*** & .42*** & -.02 & .31*** & .32*** & .34*** \\ 
 & (.049) & (.049) & (.049) & (.049) & (.049) & (.050) \\ 
AI vs. Editors & .09 & .34*** & .07 & .07 & .08 & .41*** \\ 
 & (.049) & (.049) & (.049) & (.049) & (.049) & (.049) \\ 
Google vs. Editors & -.22*** & -.08 & .09 & -.24*** & -.24*** & .06 \\ 
 & (.049) & (.049) & (.049) & (.049) & (.049) & (.049) \\ 
\midrule
\multicolumn{6}{l}{\textit{Note.} *** \textit{p} < .001, ** \textit{p} < .01, * \textit{p} < .05.}
\vspace{5pt}
    \end{tabular}
    \label{tab:effort_practice5}
\end{table}

Table \ref{tab:effort_test5} shows OLS models predicting test effort metrics from practice condition. Results show some differences: participants who practiced with AI pressed more keys but reported less subjective effort.

\begin{table}[H]
    \centering
        \caption{Test effort differences}

    \begin{tabular}{lcccccc}
\toprule
  & sqrt(Time) & log(Keystrokes) & Subjective Rating (0 - 10) & Time & Keystrokes & Keystrokes/min \\ 
\midrule
\multicolumn{7}{l}{Means} \\[2.5pt] 
\midrule
AI & 2.32 & 5.44 & 7.95 & 5.55 & 415.78 & 1.92 \\ 
 & (.056) & (.177) & (.225) & (.218) & (39.760) & (.041) \\ 
Google & 2.23 & 5.29 & 7.52 & 5.20 & 384.82 & 1.92 \\ 
 & (.056) & (.177) & (.225) & (.218) & (39.747) & (.041) \\ 
Editors & 2.29 & 5.39 & 8.00 & 5.41 & 381.98 & 1.95 \\ 
 & (.056) & (.177) & (.225) & (.218) & (39.793) & (.041) \\ 
\midrule
\multicolumn{7}{l}{Effect Sizes (d)} \\[2.5pt] 
\midrule
AI vs. Google & -.18*** & -.09 & -.22*** & -.18*** & -.09 & -.02 \\ 
 & (.049) & (.049) & (.049) & (.049) & (.049) & (.049) \\ 
AI vs. Editors & -.07 & -.03 & .02 & -.08 & -.10* & .07 \\ 
 & (.049) & (.049) & (.049) & (.049) & (.049) & (.049) \\ 
Google vs. Editors & .11* & .06 & .24*** & .11* & -.01 & .09 \\ 
 & (.049) & (.049) & (.049) & (.049) & (.049) & (.049) \\ 
\midrule
\multicolumn{6}{l}{\textit{Note.} *** \textit{p} < .001, ** \textit{p} < .01, * \textit{p} < .05.}
\vspace{5pt}    \end{tabular}
    \label{tab:effort_test5}
\end{table}

\subsection{AI practice did not create more of an illusion of mastery than getting feedback from professional editors or looking for examples online}\label{sec:future_learning5}

As reported in the main text, there were no differences between participants who practiced with AI and participants who practiced with editor feedback on how much they thought they learned, their perceived skill, and their likelihood to ask for feedback after the test. See Table \ref{tab:motivation5}

\begin{table}[H]
    \centering
        \caption{Differences in motivational variables by condition}

    \begin{tabular}{lccc}
\toprule
  & Perceived Learning & Perceived Writing Skill & Asked for Feedback \\ 
\midrule
\multicolumn{4}{l}{\textbf{Means}} \\[2.5pt] 
\midrule
{AI} & {6.56} & {6.74} & {.67} \\ 
{} & {(.227)} & {(.157)} & {(.054)} \\ 
{Google} & {6.01} & {6.59} & {.59} \\ 
{} & {(.227)} & {(.157)} & {(.058)} \\ 
{Editors} & {6.41} & {6.76} & {.66} \\ 
{} & {(.227)} & {(.156)} & {(.054)} \\ 
\midrule
\multicolumn{4}{l}{\textbf{Effect Sizes (d)}} \\[2.5pt] 
\midrule
{AI vs. Google} & {-.28***} & {-.11*} & {1.38**} \\ 
{} & {(.049)} & {(.049)} & {(.147)} \\ 
{AI vs. Editors} & {-.07} & {.02} & {1.03} \\ 
{} & {(.049)} & {(.049)} & {(.112)} \\ 
{Google vs. Editors} & {.20***} & {.12*} & {.75**} \\ 
{} & {(.049)} & {(.049)} & {(.080)} \\ 
\midrule
\multicolumn{4}{l}{\textit{Note.} *** \textit{p} < .001, ** \textit{p} < .01, * \textit{p} < .05.}
\vspace{5pt}
\end{tabular}
    \label{tab:motivation5}
\end{table}

\newpage
\subsection{Results were not moderated by individual differences}
As in Study 2, we tested whether each of the pretreatment demographic variables moderated the effects of seeing an AI example. To do this, we ran separate linear regressions in which writing skill during the test phase was regressed on condition, the pre-treatment moderator of interest, writing skill at baseline, and an interaction term between the moderator $\times$ condition. After correcting the \textit{p-}values for the interaction terms, none were significant at the .05 level, suggesting that seeing AI examples was equally effective across groups. See Table \ref{tab:interactions5}

\begin{table}[H]
    \centering
        \caption{BH-corrected \textit{p}-values for interaction terms from models predicting each outcome from condition interacted with pre-treatment variables.}

    \begin{tabular}{lrrrrrrrrrrrrrr}
\toprule
 & \multicolumn{2}{c}{\textbf{Test}} & \multicolumn{2}{c}{\textbf{Time Practice}} & \multicolumn{2}{c}{\textbf{Keys Practice}} & \multicolumn{2}{c}{\textbf{Effort Practice}} & \multicolumn{2}{c}{\textbf{Per. Learning}} & \multicolumn{2}{c}{\textbf{Per. Skill}} & \multicolumn{2}{c}{\textbf{Want Feedback}} \\ 
\cmidrule(lr){2-3} \cmidrule(lr){4-5} \cmidrule(lr){6-7} \cmidrule(lr){8-9} \cmidrule(lr){10-11} \cmidrule(lr){12-13} \cmidrule(lr){14-15}
 & \textbf{GE} & \textbf{EF} & \textbf{GE} & \textbf{EF}  & \textbf{GE} & \textbf{EF}  & \textbf{GE} & \textbf{EF}  & \textbf{GE} & \textbf{EF} &\textbf{GE} & \textbf{EF}& \textbf{GE} & \textbf{EF}  \\ 
\midrule
\multicolumn{15}{l}{} \\[2.5pt] 
\midrule
Pretest & 0.805 & 0.950 & 0.990 & 0.877 & 0.990 & 0.990 & 0.805 & 0.997 & 0.919 & 0.930 & 0.990 & 0.990 & 0.990 & 0.690 \\ 
Year of Birth & 0.996 & 0.402 & 0.990 & 0.748 & 0.877 & 0.894 & 0.798 & 0.852 & 0.765 & 0.708 & 0.936 & 0.695 & 0.990 & 0.789 \\ 
Writing Skill & 0.971 & 0.991 & 0.889 & 0.805 & 0.877 & 0.894 & 0.990 & 0.985 & 0.878 & 0.572 & 0.789 & 0.957 & 0.877 & 0.572 \\ 
\midrule
\multicolumn{15}{l}{\textbf{Gender} vs. Female} \\[2.5pt] 
\midrule
Male & 0.695 & 0.957 & 0.877 & 0.805 & 0.899 & 0.894 & 0.891 & 0.950 & 0.981 & 0.889 & 0.990 & 0.923 & 0.314 & 0.402 \\ 
Other & 0.790 & 0.889 & 0.885 & 0.990 & 0.916 & 0.990 & 0.899 & 0.798 & 0.950 & 0.861 & 0.877 & 0.990 & 0.798 & 0.419 \\ 
\midrule
\multicolumn{15}{l}{\textbf{Race/Etnicity}} \\[2.5pt] 
\midrule
White & 0.267 & 0.402 & 0.402 & 0.542 & 0.950 & 0.877 & 0.878 & 0.887 & 0.862 & 0.877 & 0.805 & 0.990 & 0.852 & 0.449 \\ 
Black & 0.748 & 0.922 & 0.748 & 0.990 & 0.891 & 0.877 & 0.442 & 0.211 & 0.189 & 0.533 & 0.140 & 0.427 & 0.990 & 0.990 \\ 
Asian & 0.150 & 0.432 & 0.990 & 0.789 & 0.894 & 0.605 & 0.519 & 0.805 & 0.937 & 0.899 & 0.990 & 0.805 & 0.805 & 0.378 \\ 
Latino & 0.189 & 0.789 & 0.402 & 0.889 & 0.318 & 0.894 & 0.981 & 0.605 & 0.899 & 0.805 & 0.852 & 0.789 & 0.990 & 0.990 \\ 
Other & 0.894 & 0.798 & 0.877 & 0.990 & 0.990 & 0.805 & 0.959 & 0.660 & 0.805 & 0.701 & 0.877 & 0.427 & 0.990 & 0.877 \\ 
\midrule
\multicolumn{15}{l}{\textbf{Education}} \\[2.5pt] 
\midrule
High school & 0.990 & 0.990 & 0.882 & 0.877 & 0.990 & 0.874 & 0.990 & 0.419 & 0.998 & 0.189 & 0.140 & 0.166 & 0.990 & 0.990 \\ 
Some college & 0.990 & 0.990 & 0.877 & 0.899 & 0.959 & 0.877 & 0.990 & 0.333 & 0.990 & 0.140 & 0.143 & 0.186 & 0.990 & 0.990 \\ 
Associate  & 0.990 & 0.990 & 0.899 & 0.690 & 0.990 & 0.798 & 0.990 & 0.542 & 0.990 & 0.140 & 0.144 & 0.166 & 0.990 & 0.990 \\ 
Bachelor's  & 0.990 & 0.990 & 0.877 & 0.878 & 0.899 & 0.877 & 0.990 & 0.438 & 0.990 & 0.143 & 0.140 & 0.172 & 0.990 & 0.990\\ 
Master's & 0.990 & 0.981 & 0.889 & 0.877 & 0.957 & 0.852 & 0.973 & 0.605 & 0.990 & 0.143 & 0.140 & 0.189 & 0.990 & 0.990 \\ 
Doctoral/Professional & 0.990 & 0.894 & 0.894 & 0.805 & 0.990 & 0.789 & 0.990 & 0.402 & 0.927 & 0.140 & 0.140 & 0.140 & 0.990 & 0.990 \\ 
\midrule
\multicolumn{15}{l}{\textbf{Motivation} (vs. Not at all)} \\[2.5pt] 
\midrule
Hardly motivated & 0.990 & 0.990 & 0.611 & 0.768 & 0.877 & 0.899 & 0.690 & 0.889 & 0.901 & 0.885 & 0.577 & 0.990 & 0.798 & 0.990 \\ 
Somewhat motivated & 0.996 & 0.957 & 0.605 & 0.805 & 0.732 & 0.854 & 0.278 & 0.542 & 0.838 & 0.990 & 0.877 & 0.877 & 0.877 & 0.934 \\ 
Very motivated & 0.990 & 0.990 & 0.419 & 0.789 & 0.790 & 0.852 & 0.419 & 0.690 & 0.936 & 0.964 & 0.763 & 0.899 & 0.878 & 0.919 \\ 
Extremely motivated & 0.990 & 0.936 & 0.547 & 0.789 & 0.690 & 0.789 & 0.448 & 0.608 & 0.894 & 0.990 & 0.660 & 0.894 & 0.789 & 0.874 \\ 
\midrule
\multicolumn{15}{l}{\textbf{Experience with AI writing assistants} (vs. None)} \\[2.5pt] 
\midrule
Hardly ever & 0.990 & 0.427 & 0.789 & 0.990 & 0.267 & 0.990 & 0.990 & 0.990 & 0.990 & 0.971 & 0.990 & 0.990 & 0.789 & 0.877 \\ 
A few times per week & 0.805 & 0.877 & 0.789 & 0.990 & 0.660 & 0.990 & 0.930 & 0.990 & 0.889 & 0.444 & 0.970 & 0.990 & 0.899 & 0.877 \\ 
About once a week & 0.805 & 0.798 & 0.278 & 0.790 & 0.189 & 0.990 & 0.990 & 0.990 & 0.981 & 0.990 & 0.990 & 0.985 & 0.981 & 0.990 \\ 
Every day & 0.899 & 0.150 & 0.957 & 0.894 & 0.990 & 0.832 & 0.790 & 0.695 & 0.790 & 0.990 & 0.852 & 0.790 & 0.790 & 0.852 \\ 
\midrule
\multicolumn{15}{p{17cm}}{\textit{Note.} Models for test performance, square-root practice time, log keystrokes, subjective effort, perceived learning and perceived writing skill or OLS models. Asking to see feedback was a binary Yes/No variable, and was modelled with logistic regression. Models match the pre-registered main specification, thereby controlling for all other pre-treatment variables. Per. = Perceived}
\vspace{5pt}
\end{tabular}
    \label{tab:interactions5}
\end{table}

\newpage
\subsection{Pairwise comparisons}
\begin{figure}[H]
    \centering
    \includegraphics[width=0.5\linewidth]{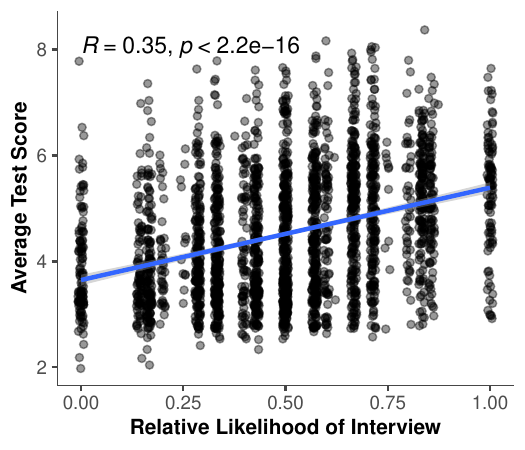}
    \caption{Correlation between test score writing quality as rated by GPT-4o and relative likelihood of being offered a hypothetical interview.}
    \label{fig:correlations4}
\end{figure}

The relative likelihood of a cover letter receiving an invitation to an interview was correlated with the GPT-rated writing quality. See Figure \ref{fig:correlations4}.

Table \ref{tab:beta4} shows the beta regressions for the relative likelihood by condition.

\begin{table}[H]
\caption{Beta regression models predicting relative likelihood of interview from condition.}
\centering

\begin{tabular}{@{\extracolsep{\fill}}lcccc}
\toprule
 & \multicolumn{2}{c}{\textbf{As preregistered}} & \multicolumn{2}{c}{\textbf{All data}} \\ 
\cmidrule(lr){2-3} \cmidrule(lr){4-5}
 & \textbf{Without controls} & \textbf{With controls} & \textbf{Without controls} & \textbf{With controls} \\ 
\midrule\addlinespace[2.5pt]
\multicolumn{5}{l}{\textbf{Means (SE)}} \\[2.5pt] 
\midrule\addlinespace[2.5pt]
AI & .50 & .51 & .50 & .50 \\ 
 & (.00) & (.01) & (.01) & (.01) \\ 
Google & .49 & .49 & .49 & .49 \\ 
 & (.00) & (.01) & (.01) & (.01) \\ 
Editors & .51 & .51 & .51 & .51 \\ 
 & (.00) & (.01) & (.01) & (.01) \\ 
\midrule\addlinespace[2.5pt]
\multicolumn{5}{l}{\textbf{Model coefficients (SE)}} \\[2.5pt] 
\midrule\addlinespace[2.5pt]
Precision ($\phi$) & 16.78*** & 7.17*** & 9.94*** & 9.94*** \\ 
 & (1.28) & (.41) & (.74) & (.74) \\ 
Symmetry (log($\nu$)) & .60*** & -.01 & -1.05*** & -1.05*** \\ 
 & (.07) & (.06) & (.09) & (.09) \\ 
\midrule\addlinespace[2.5pt]
\multicolumn{5}{l}{\textbf{Pairwise comparison (SE)}} \\[2.5pt] 
\midrule\addlinespace[2.5pt]
AI - Google & .06* & .08* & .05 & .05 \\ 
 & (.03) & (.04) & (.03) & (.03) \\ 
AI - Editors & -.01 & -.00 & -.04 & -.04 \\ 
 & (.03) & (.04) & (.03) & (.03) \\ 
Google - Editors & -.06* & -.08* & -.09** & -.09** \\ 
 & (.03) & (.04) & (.03) & (.03) \\ 
\midrule\addlinespace[2.5pt]
\multicolumn{5}{l}{\textbf{Statistics}} \\[2.5pt] 
\midrule\addlinespace[2.5pt]
N & 2575.00 & 2513.00 & 2575.00 & 2575.00 \\ 
AIC & 3697.77 & 3750.64 & 867.58 & 867.58 \\ 
BIC & 3727.04 & 3832.25 & 896.84 & 896.84 \\ 
log(Likelihood) & -1843.89 & -1861.32 & -428.79 & -428.79 \\ 
\bottomrule
\end{tabular}
    \label{tab:beta4}
\end{table}

\subsection{Forecasting and Willingness to Pay}\label{sec:mediation}
Participants predicted they would learn more from the feedback of human editors than from AI (a path = 1.05, SE = 0.22, \textit{p} \textless .001). In turn, higher predicted effectiveness predicted greater willingness to pay (b path = 0.06, SE = 0.01, \textit{p} \textless .001). Participants’ greater willingness to pay for human feedback was largely explained by their prediction that they would learn more from experienced human editors than from AI (indirect effect = 0.06, SE = 0.01, \textit{p} < .001), accounting for most of the overall tendency to pay more for human editors (total effect = 0.10, SE = 0.02, \textit{p} < .001). Once these predictions were taken into account, the remaining preference for human feedback was small and only marginally significant (c' path = 0.04, SE = 0.02, \textit{p} = .054).  See Figure \ref{fig:mediation} and Section \ref{sec:mediation} of Supplementary Information for details.

\begin{figure}[H]
    \centering
    \includegraphics[width=0.5\linewidth]{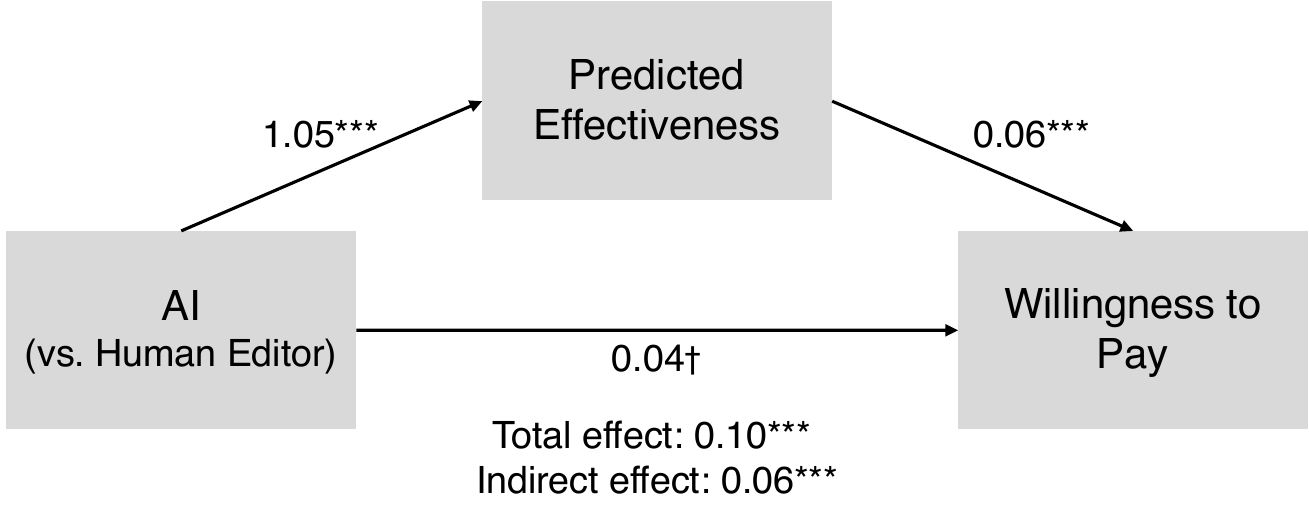}
    \caption{Predicted effectiveness mediates underinvestment in AI feedback}
    \label{fig:mediation}
\end{figure}

\section{Results Study 4}
\subsection{Randomization, Balance, and Missingness}

As in Study 2, technical issues caused small amounts of missing data. Overall, 5.64\% of data was missing in for the test phase analysis, which was not differentially missing by condition. 
There was also attrition in the follow-up sample. While most people responded, 13.45\% of recontacted participants did not respond. This attrition was not selective by condition. As shown in Table \ref{tab:s3missingness}, missingness and attrition rates were low for the main and follow-up samples and did not differ by condition.

\begin{table}[H]
    \centering
    \caption{Missingness and attrition proportions and test in Study 4.}
\begin{tabular}{lrr}
\toprule
Condition & Main Sample & Follow-up Sample \\ 
\midrule
Practice w/o AI & $4.61\%$ & $73.51\%$ \\ 
Practice w/ AI & $5.52\%$ & $70.40\%$ \\ 
See AI example & $6.77\%$ & $72.16\%$ \\ 
\midrule
Overall & $5.64\%$ & $72.04\%$ \\ 
\midrule
$\chi^2$ & $2.991$ & $1.600$ \\ 
\textit{p}-value & $0.224$ & $0.449$ \\ 
\bottomrule
\end{tabular}

    \label{tab:s3missingness}
\end{table}

Pre-treatment variables were balanced across experimental conditions, ensuring that random assignment was successful. To assess balance, we conducted a series of one-way ANOVAs for continuous variables and chi-square tests for categorical variables. Given the multiple comparisons, we applied the Benjamini-Hochberg (BH) procedure to control the false discovery rate. All statistical tests confirmed that none of the pre-treatment variables differed significantly across conditions. See Table \ref{tab:s3randomization}.

\begin{table}[H]
    \centering
\caption{Randomization checks for pre-treatment variables. \textit{p}-values are BH corrected. SMD = Standardized Mean Difference.}

\begin{tabular}{lrrrrrrr}
\toprule
\multicolumn{1}{l}{} & Overall & Practice w/o AI & Practice w/ AI & See AI example & \textit{p} & SMD \\ 
\midrule
\textit{\textbf{n}} & 2003 & 672 & 652 & 679 &  &   \\ 
\textbf{Age (mean (SD))} & 37.89 (12.63) & 37.77 (12.37) & 37.87 (12.85) & 38.03 (12.69) & .997  &  0.014 \\ 
\textbf{Gender (\%)} &  &   &   &   & .822 &   0.055 \\ 
\hspace{1em}   Female & 1056 (52.7) & 341 (50.7) & 350 (53.7) & 365 (53.8) &  &  \\ 
\hspace{1em}   Male & 923 (46.1) & 321 (47.8) & 296 (45.4) & 306 (45.1) &  &  \\ 
\hspace{1em}   Other & 24 (1.2) & 10 (1.5) & 6 (0.9) & 8 (1.2) &  &  \\ 
\textbf{Race/Ethnicity (\%)} &  &   &   &   &  &  \\ 
\hspace{1em}White = 1 & 1287 (64.3) & 419 (62.4) & 430 (66.0) & 438 (64.5) & .655 & 0.050 \\ 
\hspace{1em}Black = 1 & 484 (24.2) & 184 (27.4) & 144 (22.1) & 156 (23.0) & .324 & 0.082 \\ 
\hspace{1em}Asian = 1 & 127 (6.3) & 37 (5.5) & 43 (6.6) & 47 (6.9) & .715 & 0.039 \\ 
\hspace{1em}Latino = 1 & 163 (8.1) & 55 (8.2) & 46 (7.1) & 62 (9.1) & .655 & 0.051 \\
\hspace{1em}Other = 1 & 3 (0.1) & 1 (0.1) & 2 (0.3) & 0 (0.0) & .655 & 0.055 \\ 
\textbf{Education Level (\%)} &  &   &   &   & .655 & 0.152 \\ 
\hspace{1em}Less than high school degree & 10 (0.5) & 3 (0.4) & 3 (0.5) & 4 (0.6) & &  \\ 
\hspace{1em}High school graduate & 205 (10.2) & 74 (11.0) & 61 (9.4) & 70 (10.3) & &  \\ 
\hspace{1em}Some college, no degree & 305 (15.2) & 104 (15.5) & 110 (16.9) & 91 (13.4) & &  \\ 
\hspace{1em}Associate degree & 169 (8.4) & 68 (10.1) & 45 (6.9) & 56 (8.2) & &  \\ 
\hspace{1em}Bachelor's degree & 850 (42.4) & 255 (37.9) & 290 (44.5) & 305 (44.9) & &  \\ 
\hspace{1em}Master's degree & 401 (20.0) & 144 (21.4) & 126 (19.3) & 131 (19.3) & &  \\ 
\hspace{1em}Doctoral degree (PhD) & 36 (1.8) & 14 (2.1) & 11 (1.7) & 11 (1.6) & &  \\ 
\hspace{1em}Professional degree (JD, MD) & 27 (1.3) & 10 (1.5) & 6 (0.9) & 11 (1.6) & &  \\ 
\textbf{Writing Skill (mean (SD))} & 6.60 (1.70) & 6.63 (1.67) & 6.71 (1.69) & 6.46 (1.73) & .228 & 0.100 \\ 
\textbf{Motivation (\%)} &  &   &   &   & .997 &  0.042 \\ 
\hspace{1em}Not at all motivated & 28 (1.4) & 9 (1.3) & 10 (1.5) & 9 (1.3) &  &  \\ 
\hspace{1em}Hardly motivated & 154 (7.7) & 50 (7.4) & 53 (8.1) & 51 (7.5) &  &  \\ 
\hspace{1em}Somewhat motivated & 639 (31.9) & 221 (32.9) & 202 (31.0) & 216 (31.8) &  &  \\ 
\hspace{1em}Very motivated & 762 (38.0) & 249 (37.1) & 249 (38.2) & 264 (38.9) &  &  \\ 
\hspace{1em}Extremely motivated & 420 (21.0) & 143 (21.3) & 138 (21.2) & 139 (20.5) &  &  \\ 
\textbf{Experience with AI (\%)} &  &   &   &   & .655 & 0.103 \\ 
\hspace{1em}Never used AI writing assistant & 351 (17.5) & 128 (19.0) & 105 (16.1) & 118 (17.4) & &  \\ 
\hspace{1em}Tried AI but hardly use & 807 (40.3) & 267 (39.7) & 269 (41.3) & 271 (39.9) & &  \\ 
\hspace{1em}Use AI a few times per week & 375 (18.7) & 108 (16.1) & 133 (20.4) & 134 (19.7) & &  \\ 
\hspace{1em}Use AI about once a week & 343 (17.1) & 127 (18.9) & 102 (15.6) & 114 (16.8) & &  \\ 
\hspace{1em}Use AI every day & 127 (6.3) & 42 (6.2) & 43 (6.6) & 42 (6.2) & &  \\ 
\textbf{Pretest Writing Skill (mean (SD))} & 4.21 (0.88) & 4.23 (0.90) & 4.24 (0.90) & 4.17 (0.84) & .655 & 0.051 \\ 
\bottomrule
\end{tabular}
\label{tab:s3randomization}
\end{table}

\subsection{AI examples improve writing skill}

The AI tool improved performance while participants used it. Table \ref{tab:s3_practice} shows means and standardized differences for different measures of writing skill during the practice phase. The robustness checks included after the main specification, show that results are similar when using a different language model (Column 2), when not including control variables (Column 3), when excluding participants who admitted to cheating in the test phase (Column 4), for the subset of non-attriting participants to the follow-up phase (Column 5), and for each of the 5 principles separately (Columns 6 - 10).

\begin{figure}[H]
    \centering
    \includegraphics[width=0.75\linewidth]{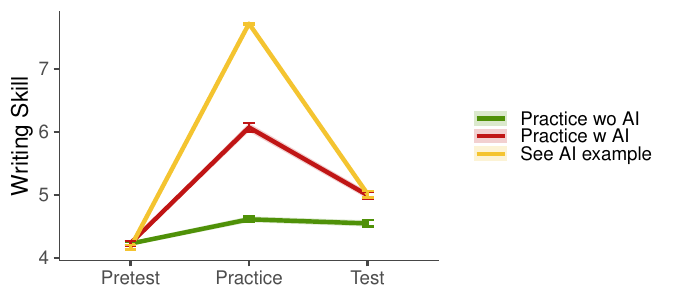}
    \caption{Participants who had practiced with the AI tool outperformed those who had practiced without it and those who had not practiced at all.
    Error bars represent means $\pm$ 1 SE.
    (\textit{N} = 2,003).}
    \label{fig:s3sup}
\end{figure}

During the test phase, when participants had to rewrite a cover letter without the help of the AI tool, participants who simply had seen an AI example outperformed participants who had practiced without the AI tool, and perfomed comparably to those who had practiced with the AI tool. Replicating Study 2, participants who had practiced with the AI tool performed better than those who had practiced  without it. Again, the learning gains are robust to different specifications, subsamples, and measures or writing quality. See Table \ref{tab:s3_test}.
For participants assigned to practice with the AI tool, The quality of AI rewrites did not correlate with participants' final submissions, \textit{r} = .06, \textit{p} = .25.

\begin{table}[H]
    \centering
    \caption{Differences in writing quality by condition in the practice phase}
   \begin{tabular}{lcccccccccc}
\toprule
  & GPT-4o & Claude & Ex. Controls & Ex. Cheaters & Follow-up & LM & ER & EN & F & ER \\ 
\midrule
\multicolumn{11}{l}{\textbf{Means --- (SE)}} \\ 
\midrule
Practice w/o AI & 4.72 & 4.96 & 4.62 & 4.61 & 4.64 & 4.33 & 6.61 & 5.43 & 2.93 & 4.31 \\ 
 & (.374) & (.379) & (.048) & (.048) & (.651) & (.396) & (.265) & (.438) & (.795) & (.466) \\ 
Practice w/ AI & 6.19 & 6.36 & 6.08 & 6.08 & 5.98 & 5.53 & 7.44 & 6.60 & 5.64 & 5.74 \\ 
 & (.373) & (.379) & (.048) & (.049) & (.656) & (.395) & (.265) & (.438) & (.794) & (.466) \\ 
See AI example & 7.83 & 8.04 & 7.72 & 7.72 & 7.58 & 7.23 & 8.38 & 8.18 & 8.34 & 7.02 \\ 
 & (.375) & (.380) & (.048) & (.049) & (.656) & (.397) & (.266) & (.439) & (.797) & (.467) \\ 
\midrule
\multicolumn{11}{l}{\textbf{Effect Sizes (d) --- (SE)}} \\ 
\midrule
Practice w/o AI vs. Practice w/ AI & 1.22*** & 1.15*** & 1.19*** & 1.21*** & 1.14*** & .95*** & .96*** & .83*** & 1.06*** & .95*** \\ 
 & (.060) & (.059) & (.059) & (.060) & (.112) & (.059) & (.059) & (.058) & (.059) & (.059) \\ 
Practice w/o AI vs. See AI example & 2.58*** & 2.52*** & 2.54*** & 2.55*** & 2.50*** & 2.28*** & 2.07*** & 1.95*** & 2.12*** & 1.81*** \\ 
 & (.070) & (.070) & (.069) & (.070) & (.132) & (.067) & (.066) & (.065) & (.066) & (.063) \\ 
Practice w/ AI vs. See AI example & 1.36*** & 1.37*** & 1.35*** & 1.35*** & 1.36*** & 1.33*** & 1.10*** & 1.12*** & 1.06*** & .86*** \\ 
 & (.061) & (.061) & (.060) & (.061) & (.113) & (.060) & (.059) & (.059) & (.059) & (.058) \\ 
\midrule
\multicolumn{11}{p{18cm}}{\textit{Note.} GPT-4o is the main specification. Ex. Controls is the main specification, unadjusted for demographic and pretreatment variables, Ex. Cheaters excludes the 3\% of participants who admitted to cheating on the test phase. LM to ER are disaggregated scores for each of the five principles. LM = Less is More, ER = Easy Reading, EN = Easy Navigation, F = Formatting, ER = Easy Responding. *** \textit{p} < .001, ** \textit{p} < .01, * \textit{p} < .05.}
\vspace{5pt}
\end{tabular}
    \label{tab:s3_practice}
\end{table}

\begin{table}[H]
    \centering
        \caption{Differences in writing quality by condition in the test phase}

\begin{tabular}{lcccccccccc}
\toprule
  & GPT-4o & Claude & Ex. Controls & Ex. Cheaters & Follow-up & LM & ER & EN & F & ER \\ 
\midrule
\multicolumn{11}{l}{\textbf{Means --- (SE)}} \\ 
\midrule
Practice w/o AI & 5.39 & 5.28 & 4.55 & 4.55 & 4.71 & 4.39 & 7.01 & 6.19 & 4.26 & 5.13 \\ 
 & (.411) & (.426) & (.055) & (.055) & (.717) & (.391) & (.367) & (.479) & (.982) & (.496) \\ 
Practice w/ AI & 5.82 & 5.84 & 5.00 & 5.00 & 5.00 & 4.69 & 7.11 & 6.61 & 5.29 & 5.38 \\ 
 & (.410) & (.426) & (.056) & (.056) & (.722) & (.390) & (.366) & (.478) & (.981) & (.495) \\ 
See AI example & 5.87 & 5.95 & 5.02 & 5.03 & 5.08 & 4.66 & 7.08 & 6.78 & 5.47 & 5.36 \\ 
 & (.412) & (.427) & (.054) & (.054) & (.722) & (.392) & (.368) & (.480) & (.985) & (.497) \\ 
\midrule
\multicolumn{11}{l}{\textbf{Effect Sizes (d) --- (SE)}} \\ 
\midrule
Practice w/o AI vs. Practice w/ AI & .32*** & .41*** & .32*** & .33*** & .22* & .24*** & .09 & .28*** & .33*** & .16** \\ 
 & (.057) & (.057) & (.056) & (.057) & (.106) & (.057) & (.057) & (.057) & (.057) & (.057) \\ 
Practice w/o AI vs. See AI example & .36*** & .49*** & .34*** & .35*** & .29** & .22*** & .06 & .39*** & .38*** & .14** \\ 
 & (.056) & (.056) & (.056) & (.057) & (.106) & (.056) & (.056) & (.056) & (.056) & (.056) \\ 
Practice w/ AI vs. See AI example & .04 & .08 & .01 & .02 & .06 & -.03 & -.02 & .11* & .06 & -.01 \\ 
 & (.056) & (.056) & (.056) & (.057) & (.104) & (.056) & (.056) & (.056) & (.056) & (.056) \\ 
\midrule
\multicolumn{11}{p{18cm}}{\textit{Note.} GPT-4o is the main specification. Ex. Controls is the main specification, unadjusted for demographic and pretreatment variables, Ex. Cheaters excludes the 3\% of participants who admitted to cheating on the test phase. LM to ER are disaggregated scores for each of the five principles. LM = Less is More, ER = Easy Reading, EN = Easy Navigation, F = Formatting, ER = Easy Responding. *** \textit{p} < .001, ** \textit{p} < .01, * \textit{p} < .05.}
\vspace{5pt}\end{tabular}

    \label{tab:s3_test}
\end{table}

\subsection{Seeing AI examples was less effortful}
Table \ref{tab:effort_practice3} shows OLS models predicting practice effort metrics from practice condition. Results show that participants seeing an AI example expended considerably less effort, measured subjectively or objectively, through keystrokes or practice time, when compared both to participants who practiced with AI and without it. As in Study 2, participants who practiced with AI still expended less effort than those who practiced without it. As pre-registered, time is square-root-transformed, and keystrokes are log-transformed. Differences are slightly smaller when using untransformed variables.

\begin{table}[H]
    \centering
        \caption{Practice effort differences}

\begin{tabular}{lcccccc}
\toprule
  & sqrt(Time) & log(Keystrokes) & Subjective Rating (0 - 10) & Time & Keystrokes & Keystrokes/min \\ 
\midrule\addlinespace[2.5pt]
\multicolumn{7}{l}{\textbf{Means}} \\[2.5pt] 
\midrule\addlinespace[2.5pt]
{Practice wo AI} & {2.83} & {5.01} & {6.17} & {9.00} & {259.34} & {27.62} \\ 
{} & {(.270)} & {(.541)} & {(.642)} & {(1.495)} & {(99.161)} & {(13.213)} \\ 
{Practice w AI} & {2.71} & {4.05} & {5.89} & {8.65} & {228.45} & {21.20} \\ 
{} & {(.270)} & {(.540)} & {(.641)} & {(1.493)} & {(99.050)} & {(13.200)} \\ 
{See AI example} & {1.85} & {.81} & {5.52} & {4.99} & {24.98} & {-8.17} \\ 
{} & {(.271)} & {(.542)} & {(.643)} & {(1.499)} & {(99.392)} & {(13.246)} \\ 
\midrule\addlinespace[2.5pt]
\multicolumn{7}{l}{\textbf{Effect Sizes (d)}} \\[2.5pt] 
\midrule\addlinespace[2.5pt]
{Practice wo AI vs. Practice w AI} & {-.14*} & {-.55***} & {-.14*} & {-.07} & {-.10} & {-.15**} \\ 
{} & {(.056)} & {(.056)} & {(.057)} & {(.056)} & {(.056)} & {(.056)} \\ 
{Practice wo AI vs. See AI example} & {-1.13***} & {-2.41***} & {-.32***} & {-.83***} & {-.73***} & {-.84***} \\ 
{} & {(.059)} & {(.067)} & {(.056)} & {(.057)} & {(.056)} & {(.056)} \\ 
{Practice w AI vs. See AI example} & {-.99***} & {-1.86***} & {-.18**} & {-.76***} & {-.64***} & {-.69***} \\ 
{} & {(.058)} & {(.063)} & {(.056)} & {(.057)} & {(.056)} & {(.056)} \\ 
\midrule
\multicolumn{6}{l}{\textit{Note.} *** \textit{p} < .001, ** \textit{p} < .01, * \textit{p} < .05.}
\vspace{5pt}
\end{tabular}
    \label{tab:effort_practice3}
\end{table}

Table \ref{tab:effort_test3} shows OLS models predicting test effort metrics from practice condition. Results show some differences: participants who had seen the AI example wrote for longer during the test, and pressed more keys; however, their subjective experience of effort was not different from those who practiced with or without the AI tool.

\begin{table}[H]
    \centering
    \caption{Test effort differences}
\begin{tabular}{lcccccc}
\toprule
   & sqrt(Time) & log(Keystrokes) & Subjective Rating (0 - 10) & Time & Keystrokes & Keystrokes/min \\ 
\midrule\addlinespace[2.5pt]
\multicolumn{7}{l}{\textbf{Means}} \\[2.5pt] 
\midrule\addlinespace[2.5pt]
{Practice wo AI} & {2.45} & {5.40} & {7.05} & {6.14} & {432.01} & {68.46} \\ 
{} & {(.170)} & {(.562)} & {(.623)} & {(.655)} & {(109.856)} & {(19.117)} \\ 
{Practice w AI} & {2.50} & {5.52} & {7.19} & {6.35} & {466.90} & {72.19} \\ 
{} & {(.169)} & {(.561)} & {(.622)} & {(.654)} & {(109.733)} & {(19.090)} \\ 
{See AI example} & {2.57} & {5.86} & {7.21} & {6.63} & {517.26} & {77.84} \\ 
{} & {(.170)} & {(.563)} & {(.625)} & {(.656)} & {(110.112)} & {(19.159)} \\ 
\midrule\addlinespace[2.5pt]
\multicolumn{7}{l}{\textbf{Effect Sizes (d)}} \\[2.5pt] 
\midrule\addlinespace[2.5pt]
{Practice wo AI vs. Practice w AI} & {.09} & {.06} & {.07} & {.10} & {.10} & {.06} \\ 
{} & {(.057)} & {(.056)} & {(.057)} & {(.057)} & {(.056)} & {(.057)} \\ 
{Practice wo AI vs. See AI example} & {.22***} & {.25***} & {.08} & {.23***} & {.24***} & {.15**} \\ 
{} & {(.056)} & {(.055)} & {(.056)} & {(.056)} & {(.055)} & {(.056)} \\ 
{Practice w AI vs. See AI example} & {.13*} & {.19***} & {.01} & {.13*} & {.14*} & {.09} \\ 
{} & {(.056)} & {(.056)} & {(.056)} & {(.056)} & {(.055)} & {(.056)} \\ 

\midrule
\multicolumn{6}{l}{\textit{Note.} *** \textit{p} < .001, ** \textit{p} < .01, * \textit{p} < .05.}
\vspace{5pt}
\end{tabular}
    \label{tab:effort_test3}
\end{table}

Table \ref{tab:learning_rate3} shows OLS models predicting learning rate metrics from practice condition. Learning rate is defined as the difference between test and pretest, divided by the effort metric. It shows how many points (10 point scale) the participant improved per unit effort (e.g., per minute spent practicing). Participants who had seen an AI example improved their skill more efficiently.

\begin{table}[H]
    \centering
    \caption{Learning rate differences}

\begin{tabular}{lcccccc}
\toprule
    & sqrt(Time) & log(Keystrokes) & Subjective Rating (0 - 10) & Time & Keystrokes & Keystrokes/min \\ 
\midrule\addlinespace[2.5pt]
\multicolumn{7}{l}{\textbf{Means}} \\[2.5pt] 
\midrule\addlinespace[2.5pt]
{Practice wo AI} & {.20} & {.15} & {.18} & {.31} & {.34} & {.20} \\ 
{} & {(.155)} & {(.272)} & {(.085)} & {(.278)} & {(.167)} & {(.277)} \\ 
{Practice w AI} & {.30} & {.22} & {.25} & {.45} & {.49} & {.30} \\ 
{} & {(.155)} & {(.272)} & {(.085)} & {(.277)} & {(.167)} & {(.277)} \\ 
{See AI example} & {.51} & {.97} & {.28} & {1.08} & {.62} & {1.00} \\ 
{} & {(.155)} & {(.273)} & {(.085)} & {(.278)} & {(.168)} & {(.278)} \\ 
\midrule\addlinespace[2.5pt]
\multicolumn{7}{l}{\textbf{Effect Sizes (d)}} \\[2.5pt] 
\midrule\addlinespace[2.5pt]
{Practice wo AI vs. Practice w AI} & {.21***} & {.08} & {.25***} & {.15**} & {.27***} & {.12*} \\ 
{} & {(.057)} & {(.057)} & {(.057)} & {(.057)} & {(.057)} & {(.057)} \\ 
{Practice wo AI vs. See AI example} & {.62***} & {.94***} & {.38***} & {.86***} & {.53***} & {.90***} \\ 
{} & {(.057)} & {(.058)} & {(.056)} & {(.058)} & {(.057)} & {(.058)} \\ 
{Practice w AI vs. See AI example} & {.41***} & {.86***} & {.13*} & {.71***} & {.25***} & {.78***} \\ 
{} & {(.057)} & {(.058)} & {(.056)} & {(.057)} & {(.056)} & {(.058)} \\ 
\midrule
\multicolumn{6}{l}{\textit{Note.} *** \textit{p} < .001, ** \textit{p} < .01, * \textit{p} < .05.}
\vspace{5pt}
\end{tabular}
    \label{tab:learning_rate3}
\end{table}

\subsection{Seeing an AI example did not discourage motivation for future learning}\label{sec:future_learning3}

Table \ref{tab:motivation3} presents differences in perceived learning, perceived writing skill, and the likelihood of asking for feedback across conditions, with effect sizes and means reported for each comparison. Despite objectively learning more, participants who practiced with AI and saw an AI example perceived their learning and skill levels to be similar to those who practiced without AI and asked for feedback at comparable rates.

\begin{table}[H]
    \centering
    \caption{Differences in motivational variables by condition}

\begin{tabular}{lccc}
\toprule
  & Perceived learning & Perceived writing skill & Asked for feedback \\ 
\midrule
\multicolumn{4}{l}{\textbf{Means --- (SE)}} \\ 
\midrule
Practice w/o AI & 5.26 & 6.33 & .64 \\ 
 & (.549) & (.517) & (.670) \\ 
Practice w/ AI & 5.25 & 6.36 & .46 \\ 
 & (.549) & (.516) & (.669) \\ 
See AI example & 5.42 & 6.23 & .55 \\ 
 & (.551) & (.518) & (.671) \\ 
\midrule
\multicolumn{4}{l}{ \textbf{Effect Sizes (d)}} \\ 
\midrule
Practice w/o AI vs. Practice w/ AI & -.01 & .02 & 1.19 \\ 
 & (.057) & (.057) & (.122) \\ 
Practice w/o AI vs. See AI example & .09 & -.06 & 1.10 \\ 
 & (.056) & (.056) & (.121) \\ 
Practice w/ AI vs. See AI example & .09 & -.08 & 0.921 \\ 
 & (.056) & (.056) & (.120) \\ 
\midrule
\multicolumn{4}{l}{\textit{Note.} *** \textit{p} < .001, ** \textit{p} < .01, * \textit{p} < .05.}
\vspace{5pt}
\end{tabular}
    \label{tab:motivation3}
\end{table}

\subsection{The benefits of seeing an AI example were just as large a day later}\label{sec:persist3}

Table \ref{tab:s3_followup} shows means and standardized differences for measures of writing skill and related outcomes during the follow-up phase. The main specification demonstrates that participants who practiced with AI continued to outperform those who did not practice or practiced without AI. Robustness checks, including using a different language model (Column 2), excluding control variables (Column 3), and removing participants who admitted to cheating (Column 4) confirm the consistency of these effects. The results also hold when evaluating each of the five principles separately (Columns 5–9). These findings suggest that the benefits of practicing with AI are durable and persist even after participants stop using the tool.

\begin{table}[H]
    \centering
        \caption{Differences in writing quality by condition in the follow-up phase}

\begin{tabular}{lccccccccc}
\toprule
  & GPT-4o & Claude & Ex. Controls & Ex. Cheaters & LM & ER & EN & F & ER \\ 
\midrule
\multicolumn{10}{l}{\textbf{Means --- (SE)}} \\ 
\midrule
Practice w/o AI & 4.95 & 5.10 & 4.87 & 4.88 & 5.32 & 6.83 & 5.40 & 1.98 & 5.23 \\ 
 & (.776) & (.798) & (.109) & (.110) & (.750) & (.730) & (.847) & (1.829) & (.905) \\ 
Practice w/ AI & 5.37 & 5.67 & 5.34 & 5.38 & 5.57 & 6.98 & 5.76 & 2.91 & 5.61 \\ 
 & (.781) & (.804) & (.103) & (.105) & (.756) & (.735) & (.853) & (1.842) & (.912) \\ 
See AI example & 5.40 & 5.71 & 5.37 & 5.36 & 5.54 & 6.98 & 5.87 & 3.14 & 5.46 \\ 
 & (.781) & (.804) & (.103) & (.105) & (.756) & (.735) & (.854) & (1.843) & (.912) \\ 
\midrule
\multicolumn{10}{l}{\textbf{Effect Sizes (d) --- (SE)}} \\ 
\midrule
Practice w/o AI vs. Practice w/ AI & .29** & .40*** & .32** & .34** & .18 & .11 & .24* & .28** & .23* \\ 
 & (.106) & (.107) & (.104) & (.106) & (.106) & (.106) & (.106) & (.106) & (.106) \\ 
Practice w/o AI vs. See AI example & .32** & .43*** & .35*** & .33** & .16 & .11 & .31** & .35*** & .14 \\ 
 & (.106) & (.107) & (.104) & (.106) & (.106) & (.106) & (.106) & (.107) & (.106) \\ 
Practice w/ AI vs. See AI example & .02 & .03 & .02 & -.01 & -.02 & -.00 & .07 & .07 & -.09 \\ 
 & (.104) & (.104) & (.101) & (.103) & (.104) & (.104) & (.104) & (.104) & (.104) \\ 
\midrule
\multicolumn{10}{p{15cm}}{\textit{Note.} GPT-4o is the main specification. Ex. Controls is the main specification, unadjusted for demographic and pretreatment variables, Ex. Cheaters excludes the 3\% of participants who admitted to cheating on the test phase. LM to ER are disaggregated scores for each of the five principles. LM = Less is More, ER = Easy Reading, EN = Easy Navigation, F = Formatting, ER = Easy Responding. *** \textit{p} < .001, ** \textit{p} < .01, * \textit{p} < .05.}
\vspace{5pt}
\end{tabular}
    \label{tab:s3_followup}
\end{table}

The follow-up analyses pool three separate follow-up samples collected on consecutive days. Table \ref{tab:s3_followup_batches} are the results for each of these samples separately.

\begin{table}[H]
    \centering
        \caption{Differences in writing quality by condition in the follow-up phase by data collection batch}

\begin{tabular}{lcccc}
\toprule
  & Overall & Batch 1 & Batch 2 & Batch 3 \\ 
\midrule
\multicolumn{5}{l}{\textbf{Means --- (SE)}} \\ 
\midrule
Practice w/o AI & 4.95 & 6.02 & 5.47 & 4.85 \\ 
 & (.776) & (1.539) & (.492) & (.811) \\ 
Practice w/ AI & 5.37 & 6.64 & 5.57 & 5.41 \\ 
 & (.781) & (1.446) & (.483) & (.817) \\ 
See AI example & 5.40 & 6.51 & 5.88 & 5.32 \\ 
 & (.781) & (1.502) & (.472) & (.818) \\ 
\midrule
\multicolumn{5}{l}{\textbf{Effect Sizes (d) --- (SE)}} \\ 
\midrule
Practice w/o AI vs. Practice w/ AI & .29** & .43 & .07 & .40** \\ 
 & (.106) & (.387) & (.187) & (.147) \\ 
Practice w/o AI vs. See AI example & .32** & .34 & .29 & .34* \\ 
 & (.106) & (.354) & (.190) & (.149) \\ 
Practice w/ AI vs. See AI example & .02 & -.09 & .22 & -.06 \\ 
 & (.104) & (.344) & (.177) & (.145) \\ 
\midrule
\multicolumn{5}{p{10cm}}{\textit{Note.} GPT-4o is the main specification. Ex. Controls is the main specification, unadjusted for demographic and pretreatment variables, Ex. Cheaters excludes the 3\% of participants who admitted to cheating on the test phase. LM to ER are disaggregated scores for each of the five principles. LM = Less is More, ER = Easy Reading, EN = Easy Navigation, F = Formatting, ER = Easy Responding. *** \textit{p} < .001, ** \textit{p} < .01, * \textit{p} < .05.}
\vspace{5pt}
\end{tabular}
    \label{tab:s3_followup_batches}
\end{table}

As mentioned in the main text the effect of practicing with AI and seeing an AI example did not become attenuated one day later. See Table \ref{tab:followup_attenuation5}.

\begin{table}[H]
\centering
\caption{OLS model interacting condition with phase (Test vs. Follow-up) shows no attenuation of the effect of practicing with AI}
\begin{tabular}{lrrrr}
\toprule
Term & Estimate & SE & t & \textit{p}-value \\ 
\midrule\addlinespace[2.5pt]
Intercept & -8.469 & 4.502 & -1.881 & 0.060 \\ 
Condition: Practice w AI & 0.421 & 0.076 & 5.551 & 0.000 \\ 
Condition: See AI example & 0.467 & 0.075 & 6.253 & 0.000 \\ 
Condition: Practice w/ AI $\times$ Follow-up & -0.007 & 0.158 & -0.044 & 0.965 \\ 
Condition: See AI example $\times$ Follow-up & -0.006 & 0.157 & -0.036 & 0.971 \\ 
Phase: Follow-Up & 0.356 & 0.114 & 3.139 & 0.002 \\ 
Pretest score & 0.394 & 0.031 & 12.877 & 0.000 \\ 
Year of birth & 0.006 & 0.002 & 2.524 & 0.012 \\ 
Gender: Male & -0.096 & 0.055 & -1.724 & 0.085 \\ 
Gender: Other & 0.117 & 0.249 & 0.470 & 0.639 \\ 
Race/ethnicity: White & 0.215 & 0.117 & 1.846 & 0.065 \\ 
Race/ethnicity: Black & -0.140 & 0.125 & -1.125 & 0.261 \\ 
Race/ethnicity: Asian & 0.575 & 0.148 & 3.880 & 0.000 \\ 
Race/ethnicity: Latino & 0.216 & 0.123 & 1.756 & 0.079 \\ 
Race/ethnicity: Other & -0.180 & 0.250 & -0.719 & 0.472 \\ 
Education: High school graduate (high school diploma or equivalent including GED) & -0.184 & 0.397 & -0.464 & 0.642 \\ 
Education: Some college but no degree & -0.187 & 0.394 & -0.475 & 0.635 \\ 
Education: Associate degree in college (2-year) & -0.207 & 0.399 & -0.518 & 0.605 \\ 
Education: Bachelor's degree in college (4-year) & -0.121 & 0.390 & -0.310 & 0.756 \\ 
Education: Master's degree & -0.033 & 0.393 & -0.085 & 0.933 \\ 
Education: Doctoral degree (PhD) & -0.227 & 0.439 & -0.518 & 0.604 \\ 
Education: Non-PhD Professional degree (JD, MD) & 0.099 & 0.456 & 0.218 & 0.828 \\ 
Writing skill & 0.045 & 0.017 & 2.578 & 0.010 \\ 
Motivation: Hardly motivated & -0.196 & 0.258 & -0.760 & 0.447 \\ 
Motivation: Somewhat motivated & -0.142 & 0.243 & -0.582 & 0.561 \\ 
Motivation: Very motivated & -0.338 & 0.244 & -1.388 & 0.165 \\ 
Motivation: Extremely motivated & -0.430 & 0.249 & -1.731 & 0.084 \\ 
Experience: I have tried AI writing assistant(s) but hardly ever use them & 0.132 & 0.079 & 1.678 & 0.093 \\ 
Experience: I use AI writing assistant(s) a few times per week & 0.003 & 0.094 & 0.034 & 0.973 \\ 
Experience: I use AI writing assistant(s) about once a week & 0.070 & 0.095 & 0.730 & 0.465 \\ 
Experience: I use AI writing assistant(s) every day & -0.063 & 0.131 & -0.480 & 0.631 \\ 
\bottomrule
\end{tabular}
\label{tab:followup_attenuation5}
\end{table}

\subsection{Seeing AI examples was equally effective across subgroups}
As in Study 2, we tested whether each of the pretreatment demographic variables moderated the effects of seeing an AI example. To do this, we ran separate linear regressions in which writing skill during the test phase was regressed on condition, the pre-treatment moderator of interest, writing skill at baseline, and an interaction term between the moderator $\times$ condition. After correcting the \textit{p-}values for the interaction terms, none were significant at the .05 level, suggesting that seeing AI examples was equally effective across groups.

\begin{table}[H]
    \centering
    \caption{\textit{p}-values for interaction terms predicting each outcome by condition and pre-treatment variables.}
    \label{tab:interactions3}
    \begin{tabular}{@{\extracolsep{-4pt}}lcccccccccccccccc}
    \toprule
    & \multicolumn{2}{c}{\textbf{Test}} & \multicolumn{2}{c}{\textbf{Follow-Up}} & \multicolumn{2}{c}{\textbf{Time Practice}} & \multicolumn{2}{c}{\textbf{Keys Practice}} & \multicolumn{2}{c}{\textbf{Effort Practice}} & \multicolumn{2}{c}{\textbf{Per. Learning}} & \multicolumn{2}{c}{\textbf{Per. Skill}} & \multicolumn{2}{c}{\textbf{Want Feedback}} \\ 
    \cmidrule(lr){2-3} \cmidrule(lr){4-5} \cmidrule(lr){6-7} \cmidrule(lr){8-9} \cmidrule(lr){10-11} \cmidrule(lr){12-13} \cmidrule(lr){14-15} \cmidrule(lr){16-17}
    \textbf{Level} & \textbf{PAI} & \textbf{AIE} & \textbf{PAI} & \textbf{AIE} & \textbf{PAI} & \textbf{AIE} & \textbf{PAI} & \textbf{AIE} & \textbf{PAI} & \textbf{AIE} & \textbf{PAI} & \textbf{AIE} & \textbf{PAI} & \textbf{AIE} & \textbf{PAI} & \textbf{AIE} \\ 
    \midrule
    
    \multicolumn{17}{l}{\textbf{Continuous Moderators}} \\ 
    \midrule
    Pretest & 0.565 & 0.546 & 0.708 & 0.945 & 0.987 & 0.857 & 0.987 & 0.940 & 0.987 & 0.405 & 0.987 & 0.967 & 0.987 & 0.576 & 0.565 & 0.274 \\ 
    YOB & 0.961 & 0.987 & 0.967 & 0.857 & 0.565 & 0.855 & 0.405 & 0.405 & 0.940 & 0.763 & 0.516 & 0.724 & 0.987 & 0.724 & 0.763 & 0.871 \\ 
    Writing Skill & 0.943 & 1.000 & 0.405 & 0.987 & 0.878 & 0.987 & 0.967 & 0.871 & 0.763 & 0.707 & 0.986 & 0.405 & 0.900 & 0.987 & 0.434 & 0.405 \\ 
    \midrule
    
    \multicolumn{17}{l}{\textbf{Gender}} \\ 
    \midrule
    Male & 0.816 & 0.987 & 0.987 & 0.987 & 0.987 & 0.532 & 0.793 & 0.535 & 0.565 & 0.405 & 0.450 & 0.703 & 0.724 & 0.535 & 0.707 & 0.565 \\ 
    Other & 0.446 & 0.565 & 0.900 & 0.987 & 0.405 & 0.724 & 0.842 & 0.728 & 0.426 & 0.791 & 0.655 & 0.987 & 0.987 & 0.763 & 0.793 & 0.791 \\ 
    \midrule
    
    \multicolumn{17}{l}{\textbf{Race}} \\ 
    \midrule
    White & 0.822 & 0.724 & 0.499 & 0.605 & 0.811 & 0.987 & 0.967 & 0.855 & 0.405 & 0.937 & 0.987 & 0.763 & 0.987 & 0.533 & 0.987 & 0.797 \\ 
    Black & 0.718 & 0.405 & 0.791 & 0.718 & 0.987 & 0.901 & 0.987 & 0.899 & 0.274 & 0.565 & 0.760 & 0.945 & 0.987 & 0.499 & 0.793 & 0.446 \\ 
    Asian & 0.940 & 0.766 & 0.734 & 0.532 & 0.703 & 0.405 & 0.934 & 0.605 & 0.341 & 0.565 & 0.937 & 0.987 & 0.989 & 0.280 & 0.535 & 0.565 \\ 
    Latino & 0.987 & 0.899 & 0.405 & 0.405 & 0.763 & 0.987 & 0.405 & 0.816 & 0.405 & 0.940 & 1.000 & 0.987 & 0.857 & 0.624 & 0.987 & 0.991 \\ 
    Other & 0.405 &  & 0.855 &  & 0.987 &  & 0.855 &  & 0.987 &  & 0.987 &  & 0.987 &  & 0.987 &  \\ 
    \midrule
    
    \multicolumn{17}{l}{\textbf{Education Level}} \\ 
    \midrule
    High School Graduate & 0.940 & 0.900 & 0.749 & 0.565 & 0.405 & 0.938 & 0.718 & 0.987 & 0.405 & 0.405 & 0.987 & 0.967 & 0.987 & 0.763 & 0.987 & 0.734 \\ 
    Some College & 0.987 & 0.987 & 0.791 & 0.565 & 0.405 & 0.900 & 0.707 & 0.986 & 0.405 & 0.405 & 1.000 & 0.987 & 0.987 & 0.878 & 0.987 & 0.987 \\ 
    Associate Degree & 0.900 & 0.987 & 0.791 & 0.565 & 0.447 & 0.855 & 0.707 & 0.940 & 0.446 & 0.520 & 0.987 & 0.987 & 0.989 & 0.811 & 0.987 & 0.814 \\ 
    Bachelor's Degree & 0.986 & 0.987 & 0.793 & 0.499 & 0.405 & 0.899 & 0.707 & 0.967 & 0.405 & 0.405 & 1.000 & 0.987 & 0.989 & 0.791 & 0.987 & 0.964 \\ 
    Master's Degree & 0.986 & 0.987 & 0.724 & 0.405 & 0.405 & 0.940 & 0.724 & 0.964 & 0.405 & 0.434 & 0.987 & 0.987 & 0.987 & 0.763 & 0.987 & 0.987 \\ 
    Doctoral Degree & 0.987 & 0.987 & 0.524 & 0.724 & 0.516 & 0.832 & 0.987 & 0.940 & 0.406 & 0.516 & 0.987 & 0.987 & 0.832 & 0.760 & 0.987 & 0.964 \\ 
    Professional Degree & 0.987 & 0.900 & 0.987 & 0.943 & 0.763 & 0.987 & 0.987 & 0.987 & 0.707 & 0.405 & 0.763 & 0.938 & 0.763 & 0.749 & 0.987 & 0.987 \\ 
    \midrule
    
    \multicolumn{17}{l}{\textbf{Motivation}} \\ 
    \midrule
    Hardly Motivated & 0.987 & 0.900 & 0.624 & 0.899 & 0.768 & 0.763 & 0.832 & 0.987 & 0.763 & 0.900 & 0.685 & 0.707 & 0.760 & 0.535 & 0.405 & 0.565 \\ 
    Somewhat Motivated & 0.977 & 0.535 & 0.763 & 0.763 & 0.871 & 0.585 & 0.937 & 0.987 & 0.763 & 0.791 & 0.791 & 0.763 & 0.797 & 0.707 & 0.536 & 0.763 \\ 
    Very Motivated & 0.907 & 0.705 & 0.778 & 0.703 & 0.986 & 0.405 & 0.797 & 0.987 & 0.797 & 0.708 & 0.847 & 0.857 & 0.847 & 0.752 & 0.451 & 0.749 \\ 
    Extremely Motivated & 0.855 & 0.536 & 0.724 & 0.707 & 0.763 & 0.724 & 0.763 & 0.987 & 0.987 & 0.763 & 0.987 & 0.900 & 0.763 & 0.763 & 0.535 & 0.535 \\ 
    \midrule
    
    \multicolumn{17}{l}{\textbf{Experience with AI Writing Assistants}} \\ 
    \midrule
    Hardly Ever Use Them & 0.724 & 0.763 & 0.907 & 0.791 & 0.405 & 0.900 & 0.405 & 0.987 & 0.967 & 0.446 & 0.565 & 0.987 & 0.987 & 0.724 & 0.987 & 0.405 \\ 
    Use a Few Times Per Week & 0.576 & 0.987 & 0.763 & 0.763 & 0.766 & 0.766 & 0.763 & 0.987 & 0.565 & 0.152 & 0.903 & 0.763 & 0.783 & 0.405 & 0.945 & 0.405 \\ 
    Use About Once a Week & 0.846 & 0.707 & 0.623 & 0.786 & 0.406 & 0.855 & 0.405 & 0.987 & 0.763 & 0.718 & 0.967 & 0.987 & 0.899 & 0.405 & 0.707 & 0.797 \\ 
    Use Every Day & 0.987 & 0.724 & 0.987 & 0.763 & 0.341 & 0.763 & 0.405 & 0.987 & 0.405 & 0.280 & 0.943 & 0.987 & 0.899 & 0.763 & 0.734 & 0.987 \\ 
    \midrule
    \multicolumn{17}{p{\textwidth}}{\textit{Note.} Models for test and follow-up performance, square-root practice time, log keystrokes, subjective effort, perceived learning and perceived writing skill or OLS models. Asking to see feedback was a binary Yes/No variable, and was modelled with logistic regression. Models match the pre-registered main specification, and thus control for all other pre-treatment variables. Per. = Perceived, PAI = Practice with AI, AIE = See AI example.}
\vspace{5pt}
    \end{tabular}
\end{table}

\newpage
\subsection{Pairwise Comparisons}
\begin{figure}[H]
    \centering
    \includegraphics[width=0.5\linewidth]{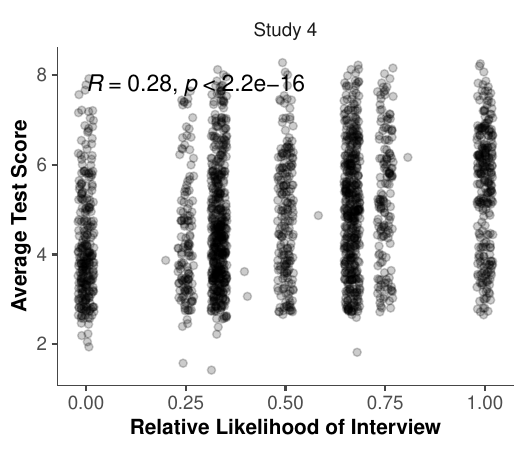}
    \caption{Correlation between test score writing quality as rated by GPT-4o and relative likelihood of being offered a hypothetical interview.}
    \label{fig:correlations5}
\end{figure}

The relative likelihood of a cover letter receiving an invitation to an interview was correlated with the GPT-rated writing quality. See Figure \ref{fig:correlations5}.

As shown in Table \ref{tab:beta5}, participants who had practiced writing cover letters with AI were more likely to be invited to a hypothetical job interview.

\begin{table}[H]
\centering
\caption{Beta regression models predicting relative likelihood of interview from condition.}
\begin{tabular}{lcc}
\toprule
 & Without controls & With controls \\ 
\midrule\addlinespace[2.5pt]
\multicolumn{3}{l}{\textbf{Means (SE)}} \\[2.5pt] 
\midrule\addlinespace[2.5pt]
Practice w AI & .50 & .51 \\ 
 & (.01) & (.01) \\ 
Practice wo AI & .48 & .48 \\ 
 & (.01) & (.01) \\ 
See AI example & .51 & .51 \\ 
 & (.01) & (.01) \\ 
\midrule\addlinespace[2.5pt]
\multicolumn{3}{l}{\textbf{Model coefficients (SE)}} \\[2.5pt] 
\midrule\addlinespace[2.5pt]
Precision ($\phi$) & 16.17*** & 7.77*** \\ 
 & (1.47) & (.51) \\ 
Symmetry (Log($\nu$))& .35*** & -.23*** \\ 
 & (.08) & (.07) \\ 
\midrule\addlinespace[2.5pt]
\multicolumn{3}{l}{\textbf{Pairwise comparison (SE)}} \\[2.5pt] 
\midrule\addlinespace[2.5pt]
Practice w AI - Practice wo AI & .08** & .09* \\ 
 & (.03) & (.04) \\ 
Practice w AI - See AI example & -.02 & -.03 \\ 
 & (.03) & (.04) \\ 
Practice wo AI - See AI example & -.10*** & -.12** \\ 
 & (.03) & (.04) \\ 
\midrule\addlinespace[2.5pt]
\multicolumn{3}{l}{\textbf{Statistics}} \\[2.5pt] 
\midrule\addlinespace[2.5pt]
N & 1934.00 & 1916.00 \\ 
AIC & 2447.07 & 2518.06 \\ 
BIC & 2474.91 & 2684.80 \\ 
log(Likelihood) & -1218.54 & -1229.03 \\ 
\bottomrule
\end{tabular}
\label{tab:beta5}
\end{table}

\end{appendices}
\end{document}